\documentclass[a4paper, twocolumn]{revtex4}
\usepackage{rotating}
\usepackage{epstopdf}
\usepackage{color}
\usepackage{float}
\usepackage{amsmath}
\usepackage{graphicx}
\usepackage{hyperref}
\usepackage{color}
\usepackage{amssymb}
\usepackage{dcolumn}
\usepackage{bm}
\usepackage{epsfig}
\usepackage{array}
\usepackage{subfigure}
\usepackage[utf8]{inputenc}
\usepackage{upgreek}
\usepackage{wasysym}
\usepackage{slashed}
\usepackage[section]{placeins}
\newcommand{\figDel}{
\begin{figure*}
\centering
\hspace{0cm}
\subfigure[noonleline][]
{\label{fig:Del1}\includegraphics[height=40mm,width=50mm]{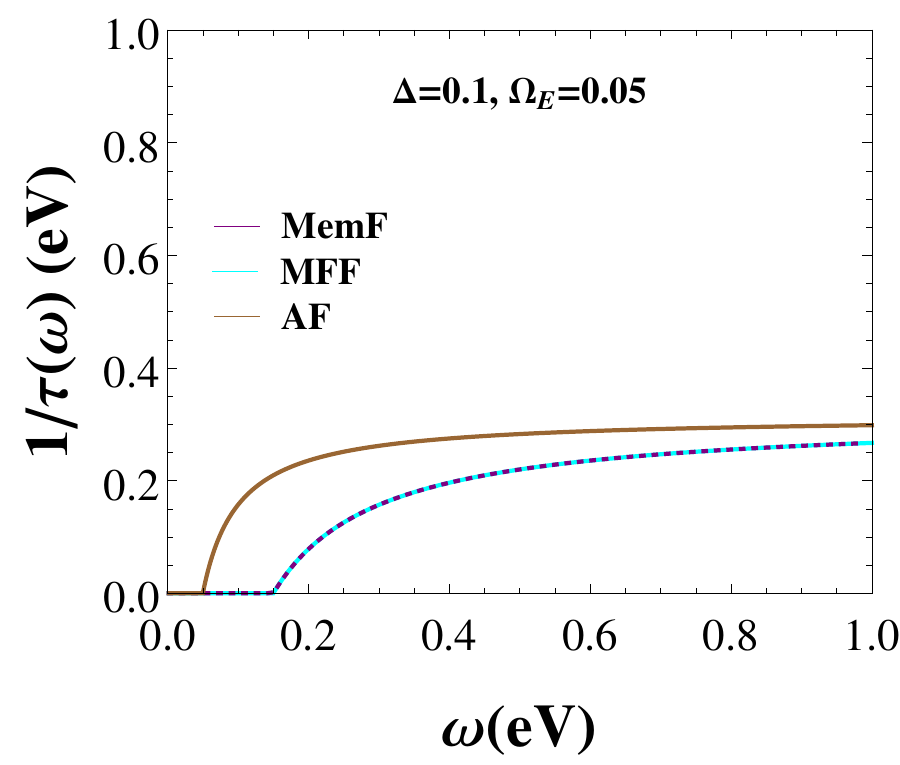}}
\hspace{0cm}
\subfigure[noonleline][]
{\label{fig:Del2}\includegraphics[height=40mm,width=50mm]{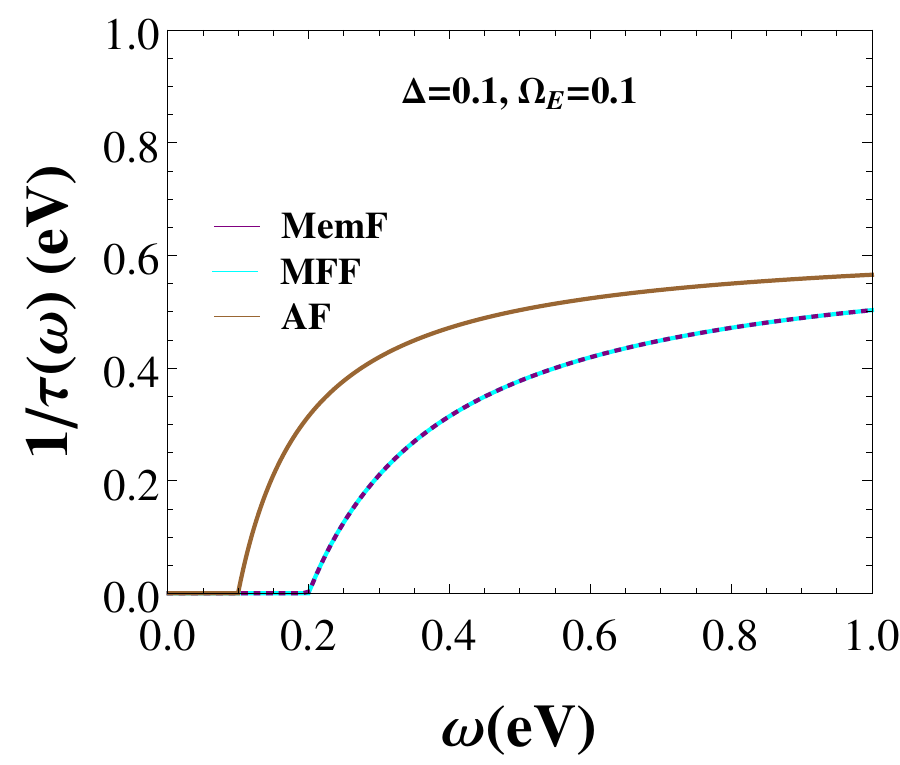}}
\subfigure[noonleline][]
{\label{fig:Del3}\includegraphics[height=40mm,width=50mm]{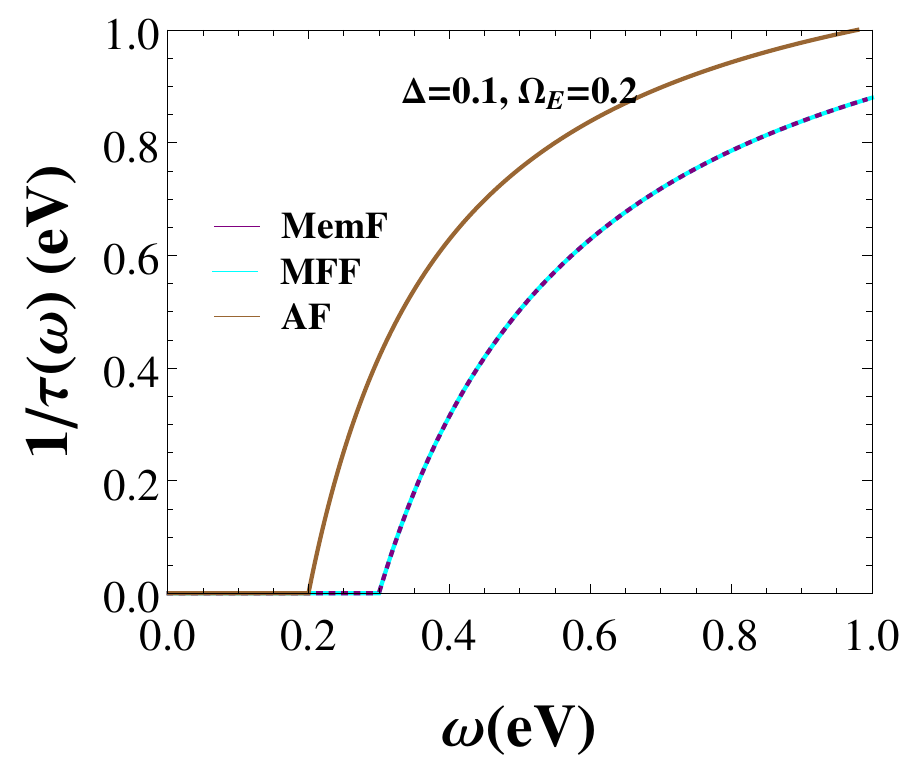}}
\caption{Scattering rate as a function of frequency from approaches (1) Allen approach (AF; Brown), (2) Mitrovi\'c-Fiorucci (MFF; Cyan), (3) Memory (MemF; Purple) using Delta function for phonon density of states. Here we have fixed the phonon peak frequency $\Omega_{E}$. (a): 0.01, (b): 0.1, (c): 0.2. For MFF and MemF, we use gap value $\Delta=0.1$eV.}
\label{fig:Del}
\end{figure*}
}
\newcommand{\figDeltemp}{
\begin{figure*}[htp]
\centering
\hspace{0cm}
\subfigure[noonleline][]
{\label{fig:Deltemp1}\includegraphics[height=40mm,width=60mm]{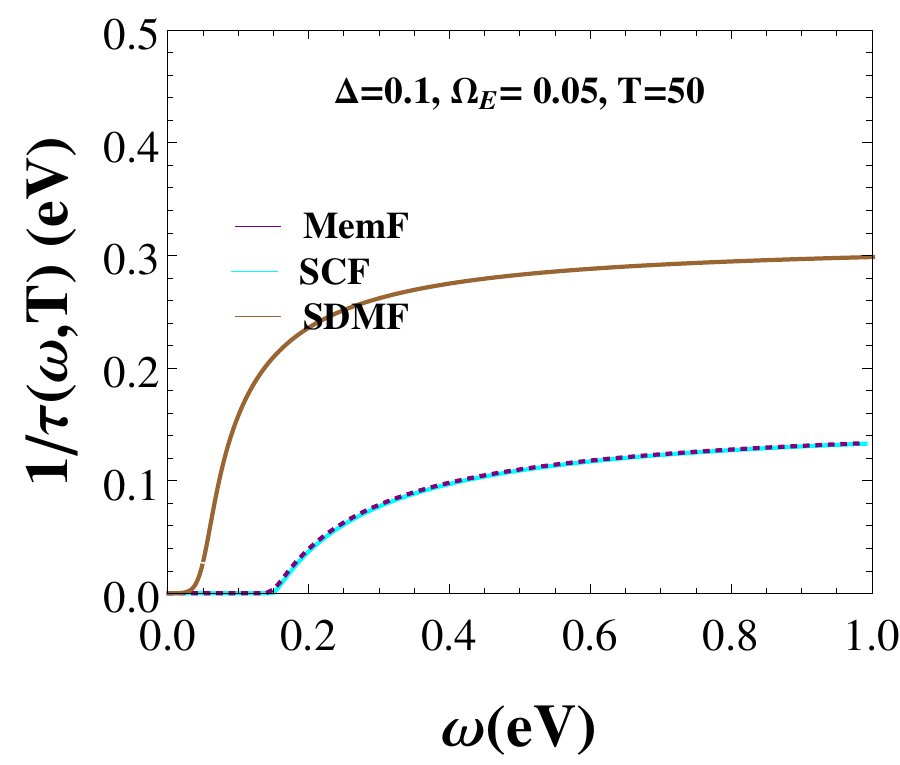}}
\hspace{0cm}
\subfigure[noonleline][]
{\label{fig:Deltemp2}\includegraphics[height=40mm,width=60mm]{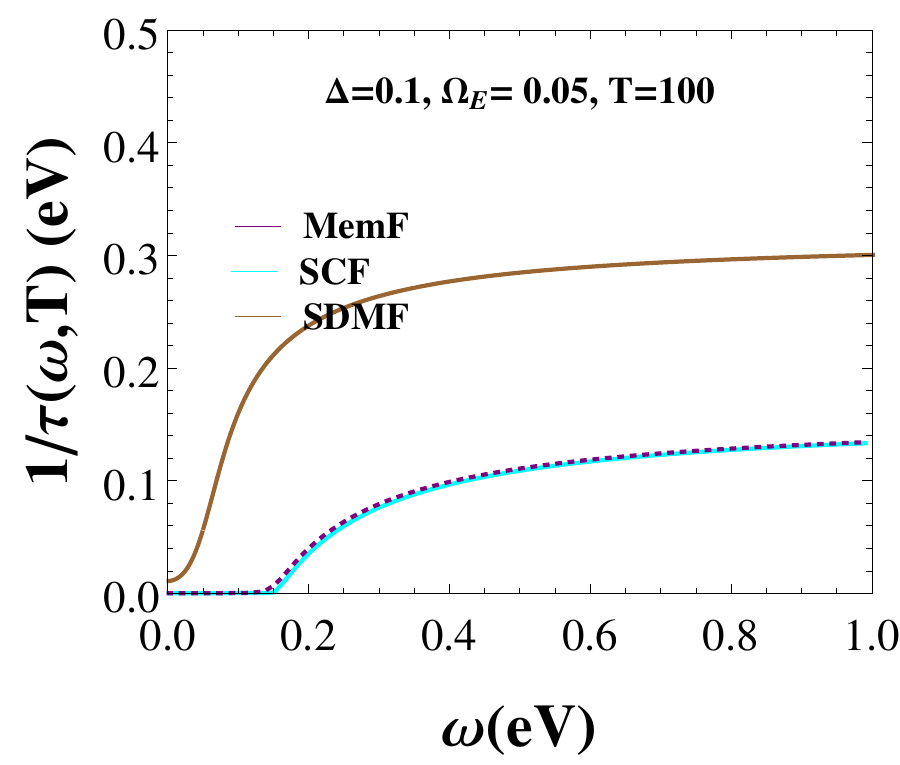}}
\subfigure[noonleline][]
{\label{fig:Deltemp3}\includegraphics[height=40mm,width=60mm]{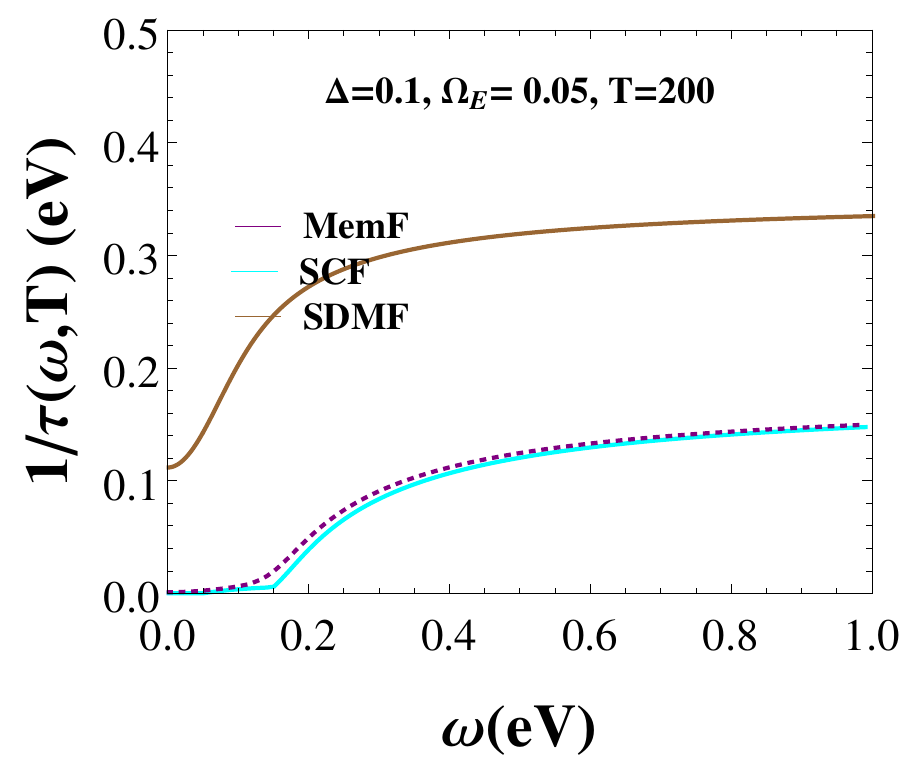}}
\subfigure[noonleline][]
{\label{fig:Deltemp4}\includegraphics[height=40mm,width=60mm]{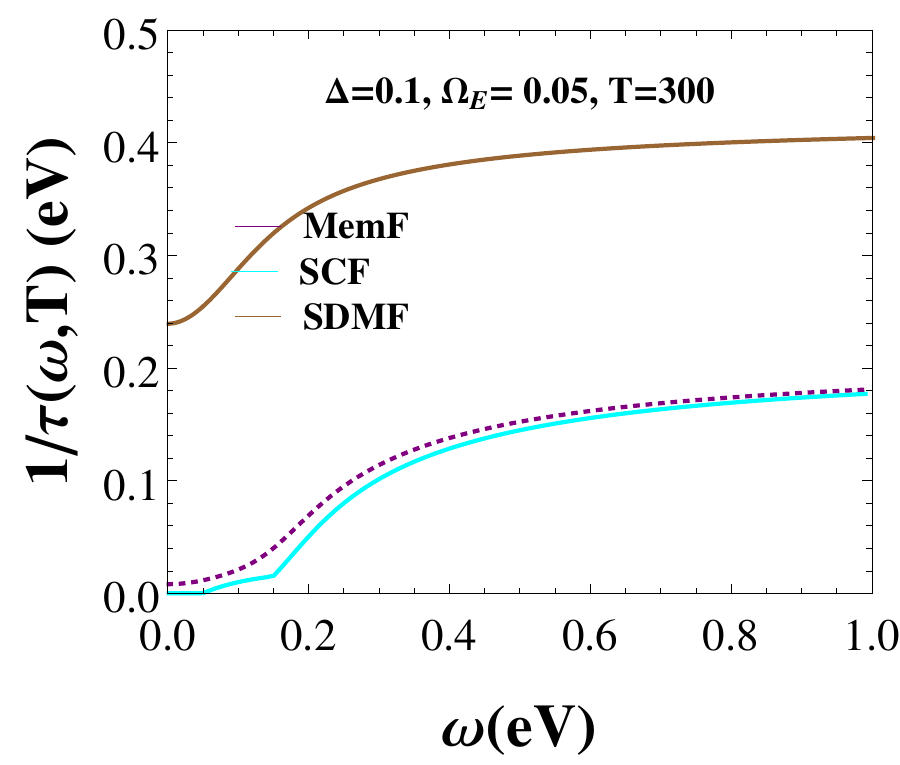}}
\caption{Temperature dependent scattering rate as a function of frequency from approaches (1) Shulga et al. (SDMF; Brown), (2) Sharapov-Carbotte (SCF; Cyan), (3) Memory (MemF; Purple) using Delta function for phonon density of states. Here we have fixed the phonon peak frequency $\Omega_{E}=0.05$ eV and consider the gap gap value $\Delta=0.1$ eV for SCF and MemF. These plots are at different temperatures such as (a): $50$, (b): $100$, (c): $200$, (d): $300$ eV.}
\label{fig:Deltemp}
\end{figure*}
}
\newcommand{\figLorent}{
\begin{figure*}[htp]
\centering
\hspace{0cm}
\subfigure[noonleline][]
{\label{fig:Loren1}\includegraphics[height=40mm,width=50mm]{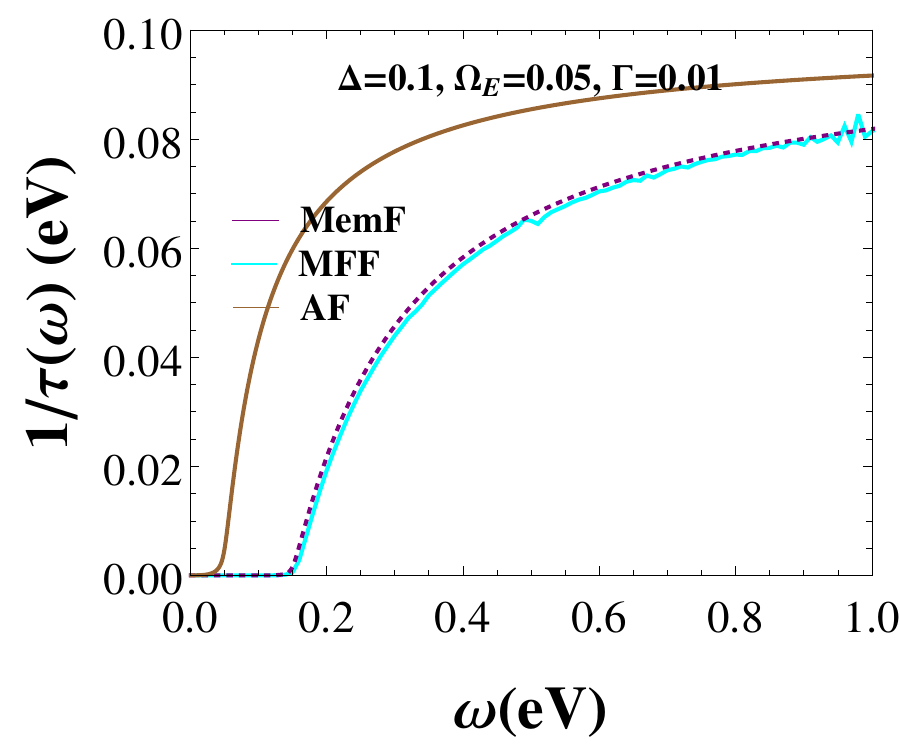}}
\hspace{0cm}
\subfigure[noonleline][]
{\label{fig:Loren2}\includegraphics[height=40mm,width=50mm]{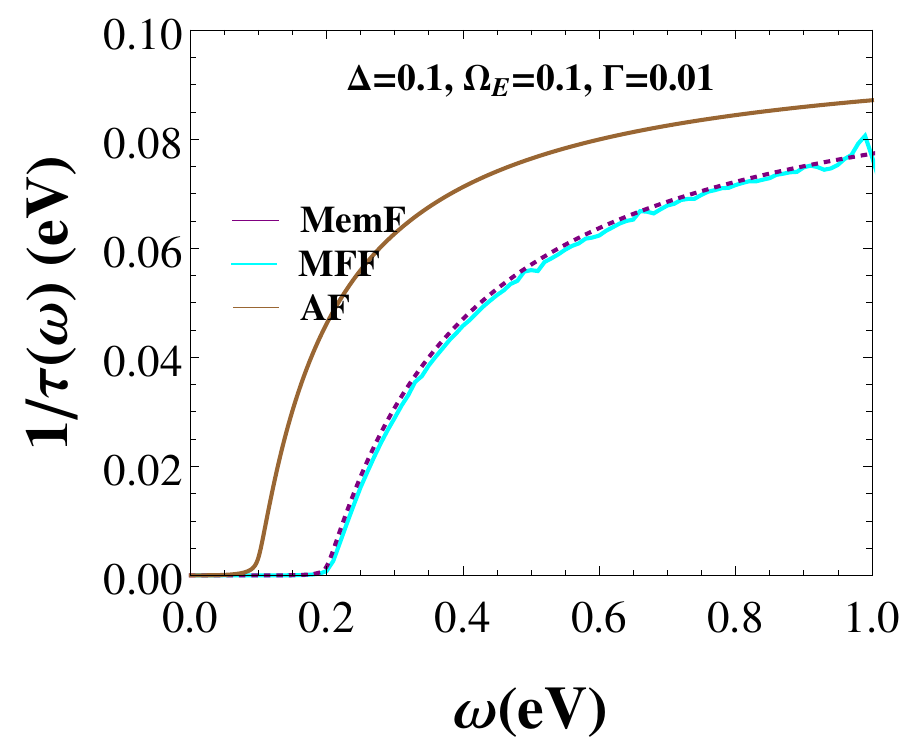}}
\subfigure[noonleline][]
{\label{fig:Loren3}\includegraphics[height=40mm,width=50mm]{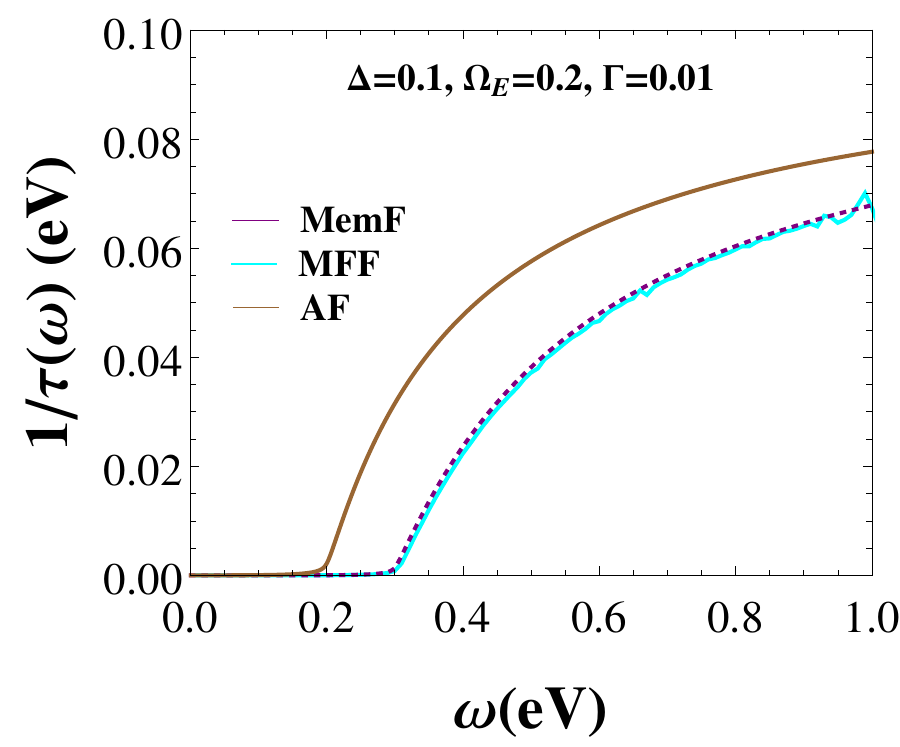}}
\subfigure[noonleline][]
{\label{fig:Loren4}\includegraphics[height=40mm,width=50mm]{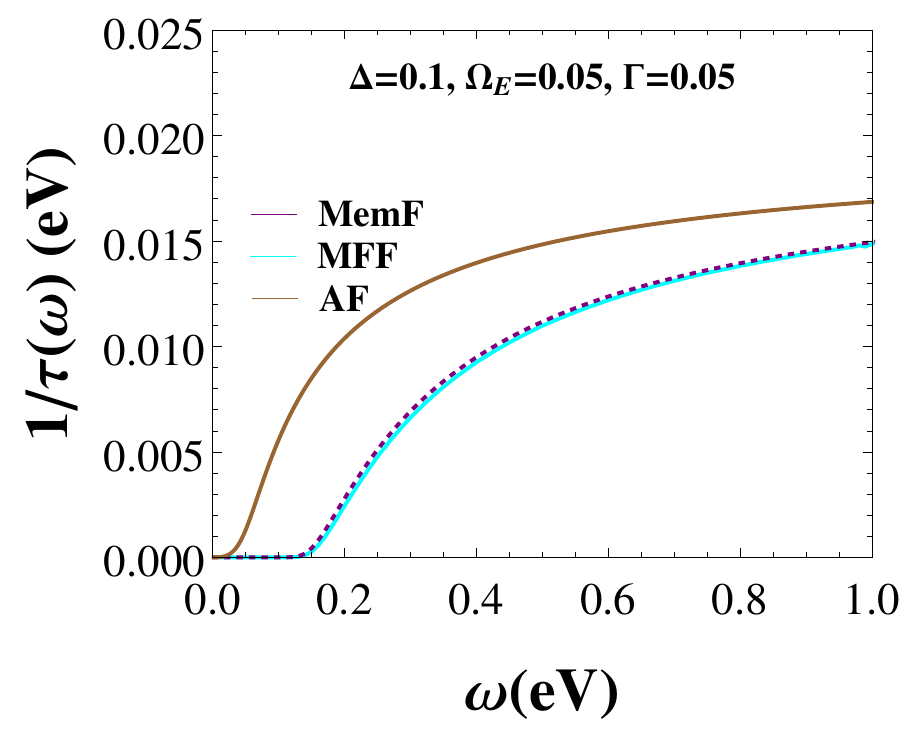}}
\subfigure[noonleline][]
{\label{fig:Loren5}\includegraphics[height=40mm,width=50mm]{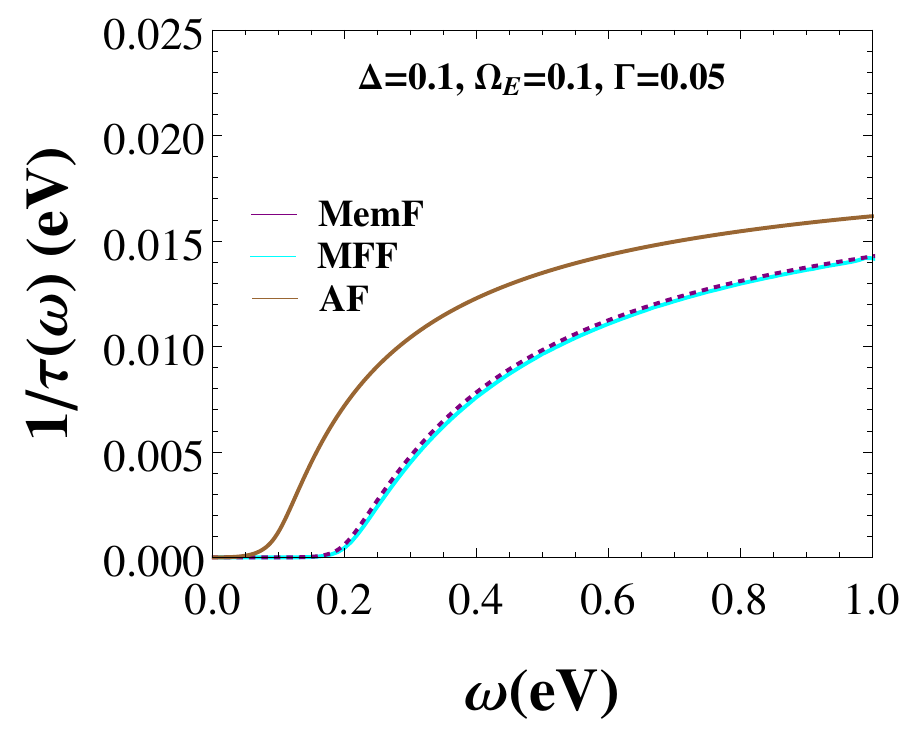}}
\subfigure[noonleline][]
{\label{fig:Loren6}\includegraphics[height=40mm,width=50mm]{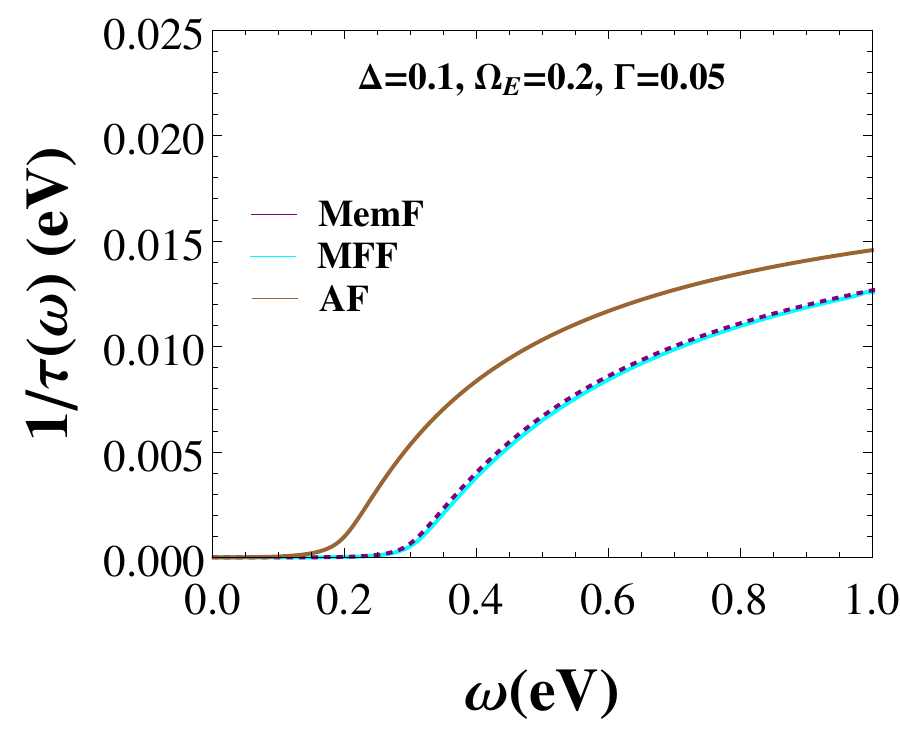}}
\caption{Scattering rate as a function of frequency from approaches (1) Allen approach (AF; Brown), (2) Mitrovi\'c-Fiorucci (MFF; Cyan), (3) Memory (MemF; Purple) using Lorentzian function for phonon density of states. Here we consider the phonon peak frequency $\Omega_{E}=0.05$, $0.01$ eV and Lorentzian width $\Gamma=0.01$, $0.05$ eV. For MFF and MemF, we have used gap value $\Delta=0.1$ eV.}
\label{fig:Lorent}
\end{figure*}
}

\newcommand{\figLorentemp}{
\begin{figure*}[htp]
\centering
\hspace{0cm}
\subfigure[noonleline][]
{\label{fig:Lorentemp1}\includegraphics[height=40mm,width=60mm]{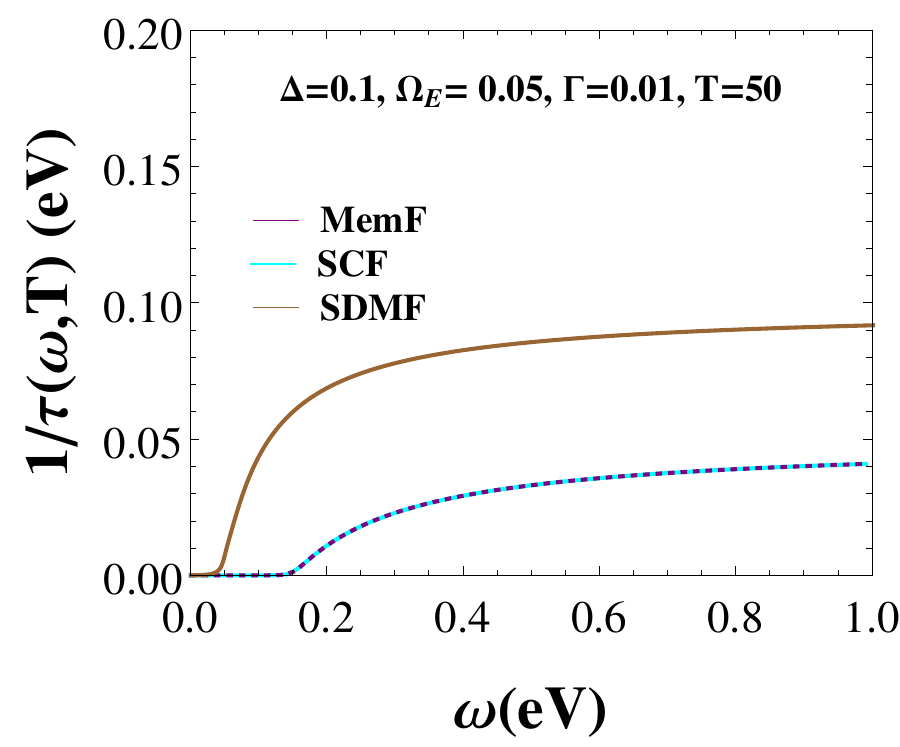}}
\hspace{0cm}
\subfigure[noonleline][]
{\label{fig:Lorentemp2}\includegraphics[height=40mm,width=60mm]{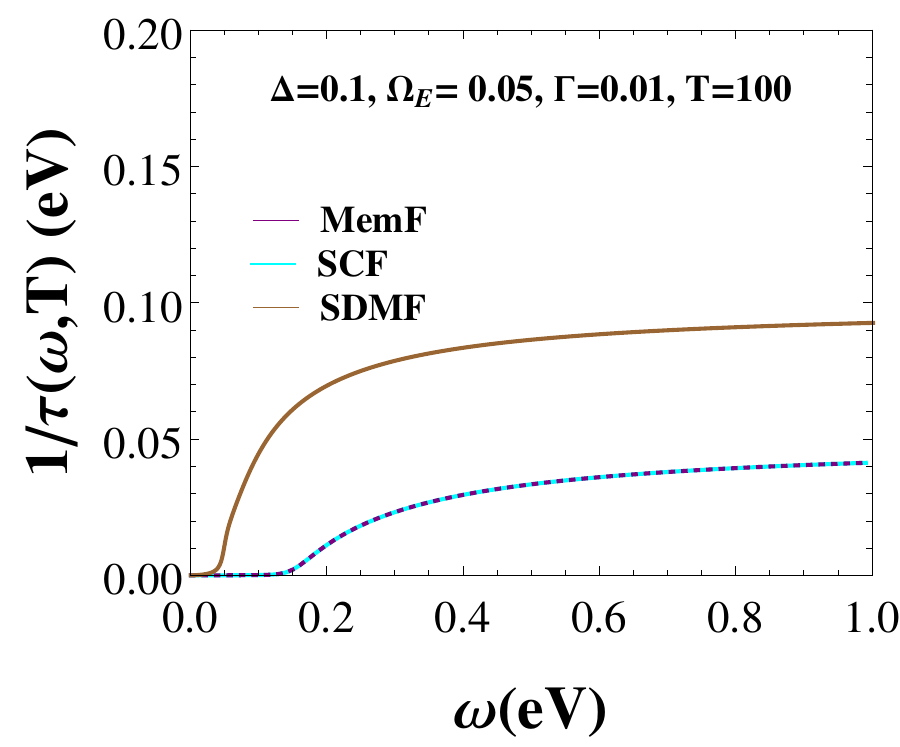}}
\subfigure[noonleline][]
{\label{fig:Lorenemp3}\includegraphics[height=40mm,width=60mm]{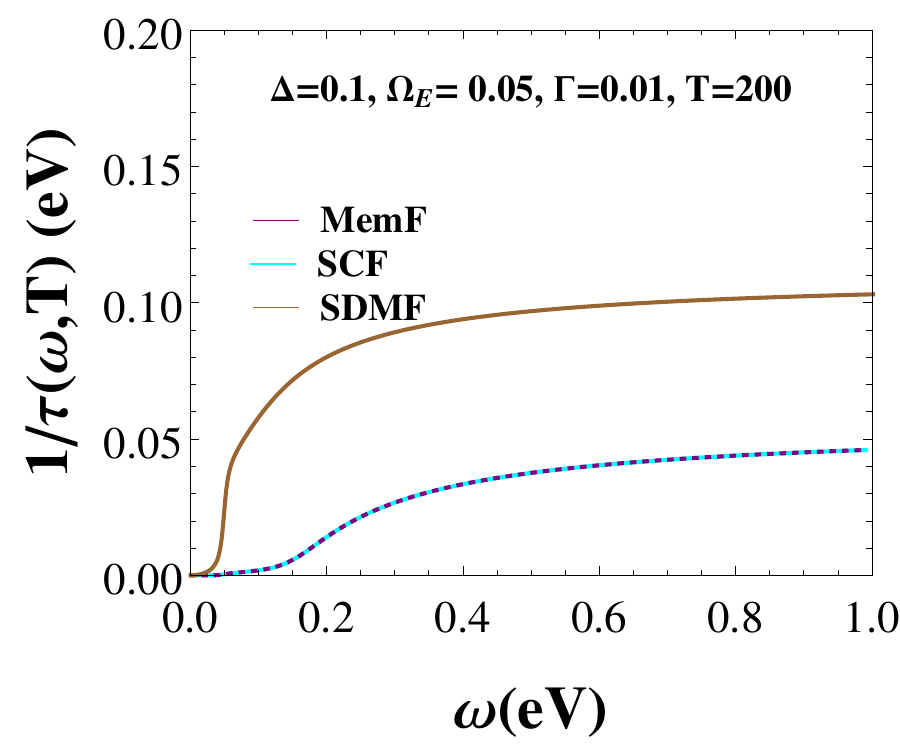}}
\subfigure[noonleline][]
{\label{fig:Lorenemp4}\includegraphics[height=40mm,width=60mm]{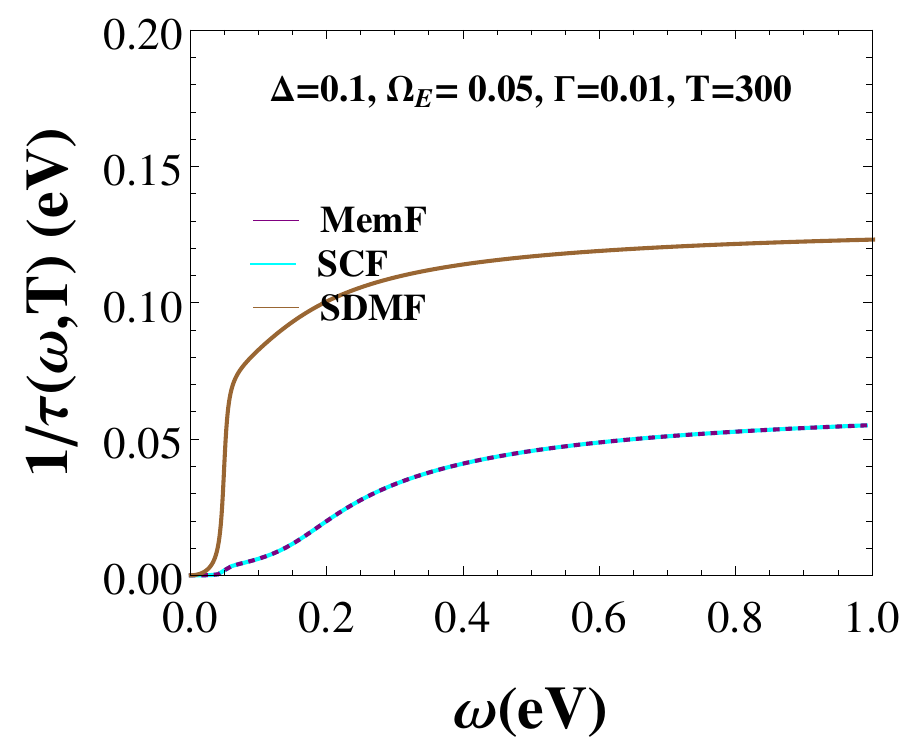}}
\caption{Temperature dependent scattering rate as a function of frequency with approaches namely Shulga et al. approach (SDMF; Brown), Sharapov-Carbotte (SCF; Cyan), Memory (MemF; Purple) using Lorentzian function for phonon density of states. Here we have fixed the phonon peak frequency $\Omega_{E}=0.05$ eV, Lorentzian width $\Gamma = 0.01$ eV and consider the gap gap value $\Delta=0.1$ eV for SCF and MemF. These plots are at different temperatures such as (a): $50$, (b): $100$, (c): $200$, (d): $300$ eV.}
\label{fig:Lorentemp}
\end{figure*}
}

\newcommand{\figdoubLorentA}{
\begin{figure*}
\centering
\hspace{0cm}
\subfigure[noonleline][]
{\label{fig:doubLoren1}\includegraphics[height=40mm,width=50mm]{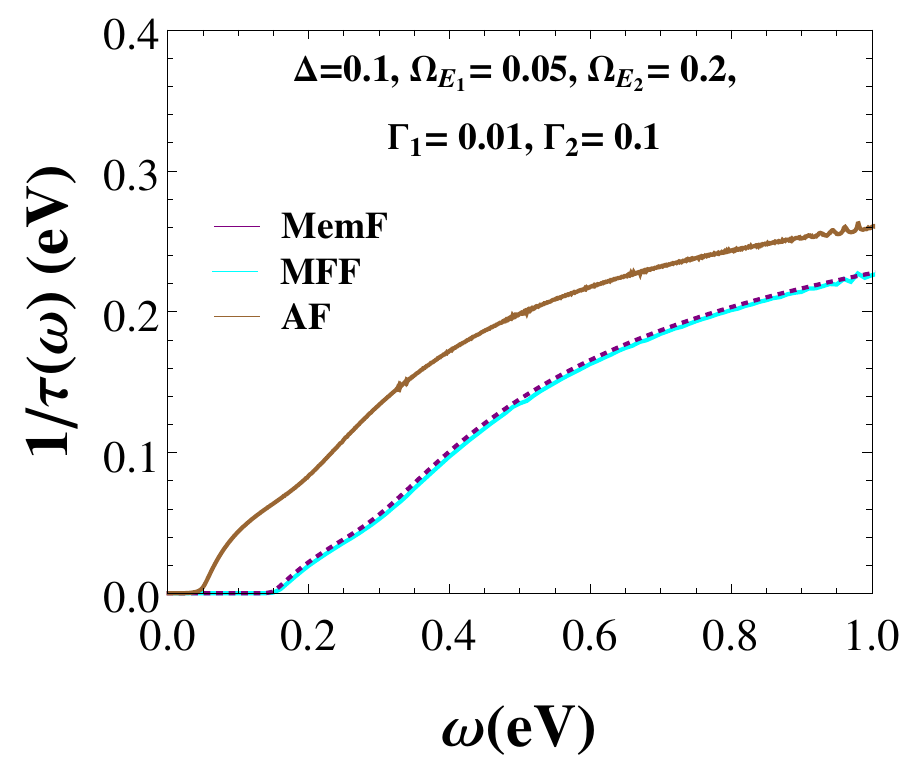}}
\hspace{0cm}
\subfigure[noonleline][]
{\label{fig:doubLoren2}\includegraphics[height=40mm,width=50mm]{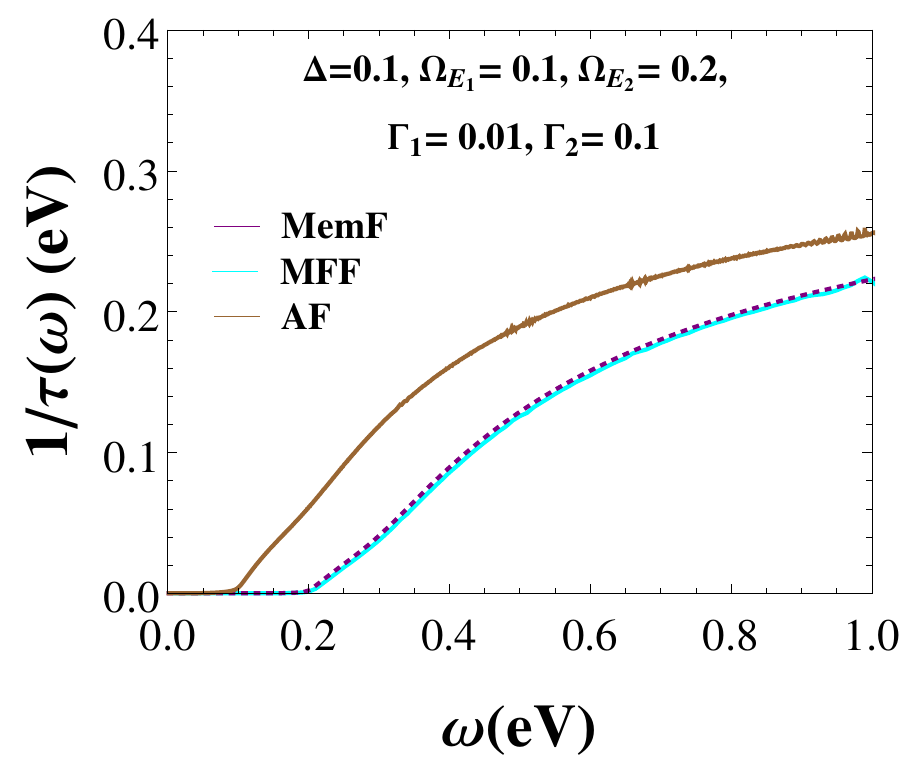}}
\subfigure[noonleline][]
{\label{fig:doubLoren3}\includegraphics[height=40mm,width=50mm]{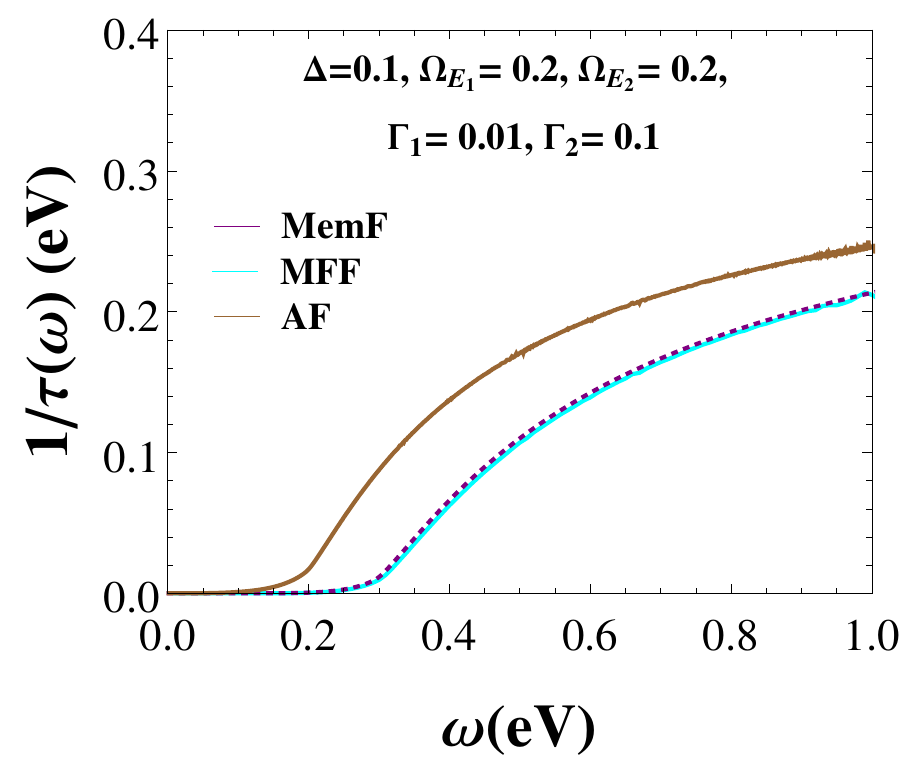}}
\subfigure[noonleline][]
{\label{fig:doubLoren4}\includegraphics[height=40mm,width=50mm]{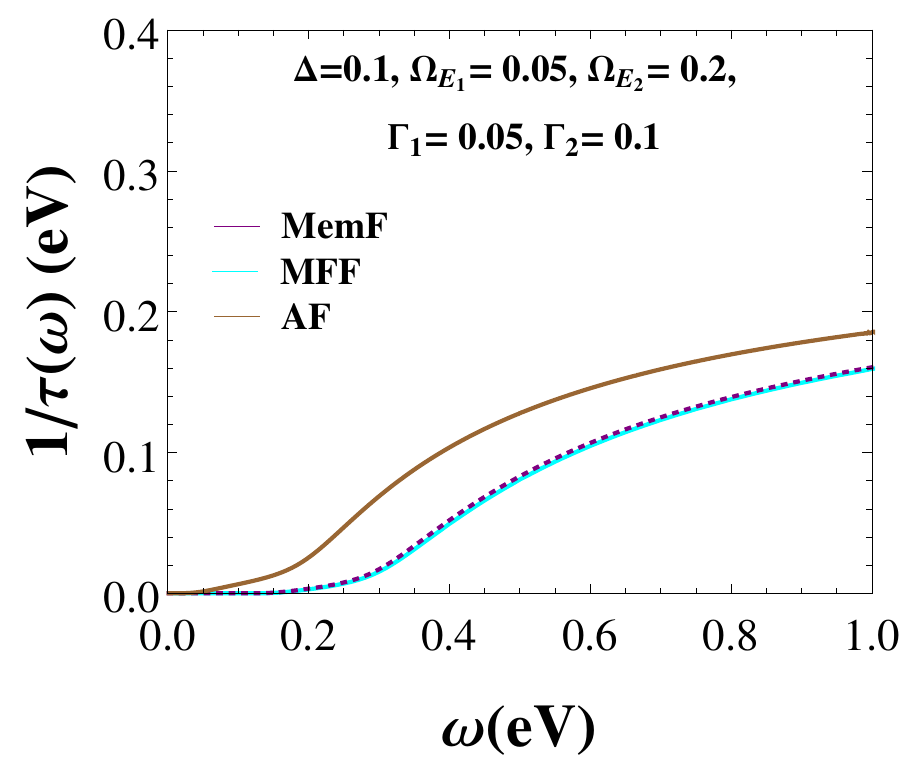}}
\subfigure[noonleline][]
{\label{fig:doubLoren5}\includegraphics[height=40mm,width=50mm]{doublelorentziandeltap1peakp05p2widthp05p1.pdf}}
\subfigure[noonleline][]
{\label{fig:doubLoren6}\includegraphics[height=40mm,width=50mm]{doublelorentziandeltap1peakp05p2widthp05p1.pdf}}
\caption{Scattering rate as a function of frequency with approaches namely Allen approach (AF; Brown), Mitrovi\'c-Fiorucci (MFF; Cyan), Memory (MemF; Purple) using Double Lorentzian function for phonon density of states. Here we consider the phonon peak frequencies $\Omega_{E_{1}}=0.05$, $0.01$, $0.2$ eV, $\Omega_{E_{2}}=0.2$ eV and Lorentzian widths $\Gamma_{1}=0.01$, $0.05$ eV, $\Gamma_{2}=0.1$ eV. For MFF and MemF, we have used gap value $\Delta=0.1$ eV.}
\label{fig:doubLorentA}
\end{figure*}
}

\newcommand{\figdoubLorentB}{
\begin{figure*}
\centering
\hspace{0cm}
\subfigure[noonleline][]
{\label{fig:doubLoren7}\includegraphics[height=40mm,width=60mm]{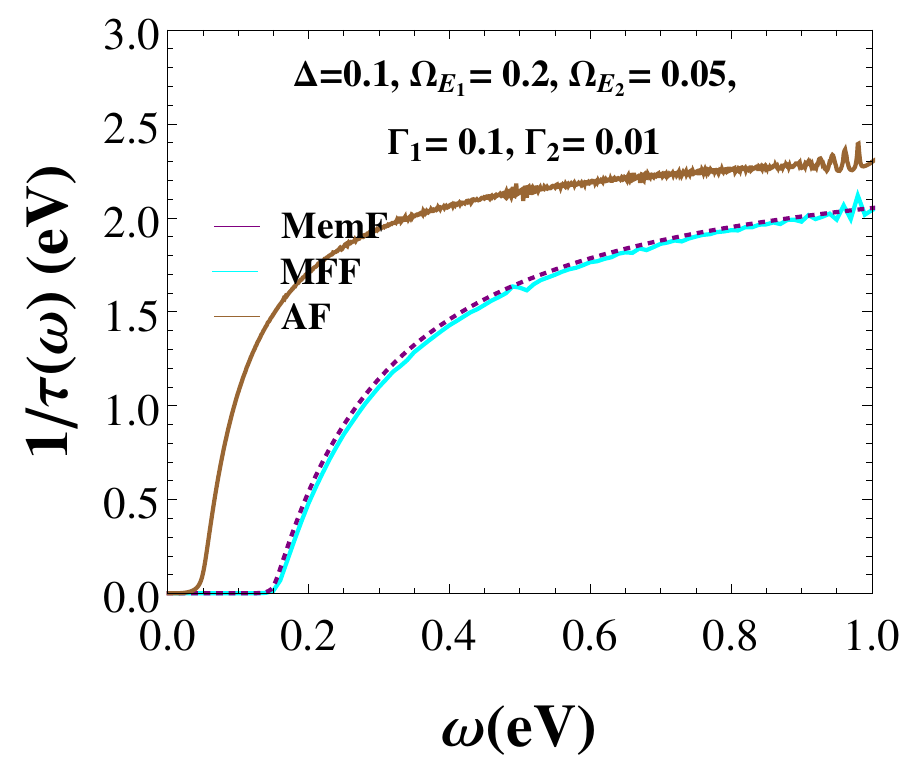}}
\hspace{0cm}
\subfigure[noonleline][]
{\label{fig:doubLoren8}\includegraphics[height=40mm,width=50mm]{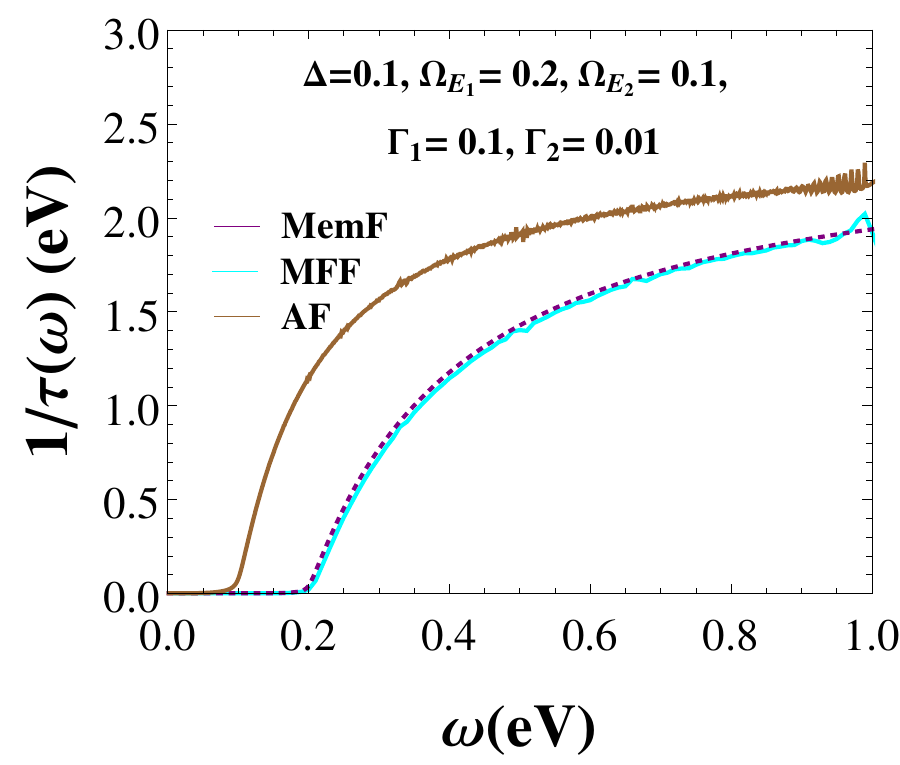}}
\subfigure[noonleline][]
{\label{fig:doubLoren9}\includegraphics[height=40mm,width=50mm]{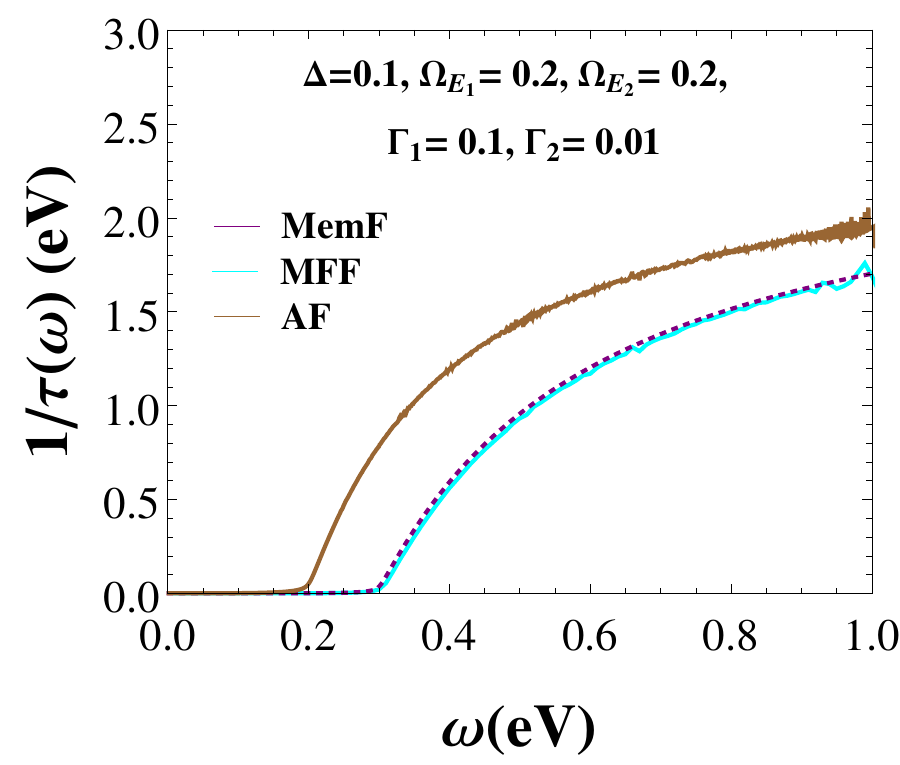}}
\subfigure[noonleline][]
{\label{fig:doubLoren10}\includegraphics[height=40mm,width=50mm]{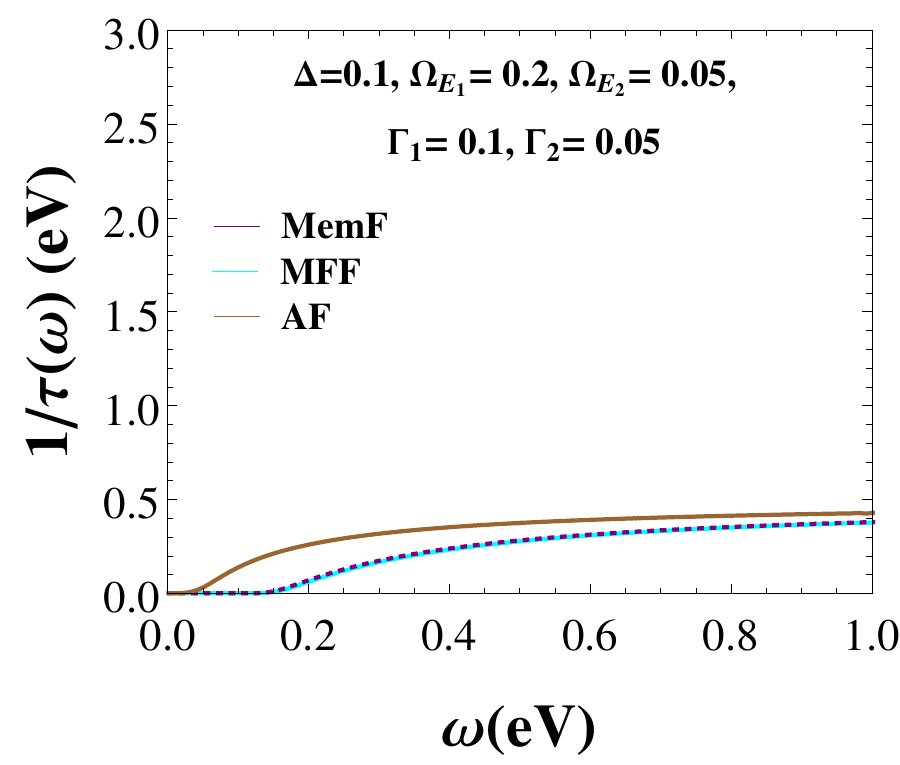}}
\subfigure[noonleline][]
{\label{fig:doubLoren11}\includegraphics[height=40mm,width=50mm]{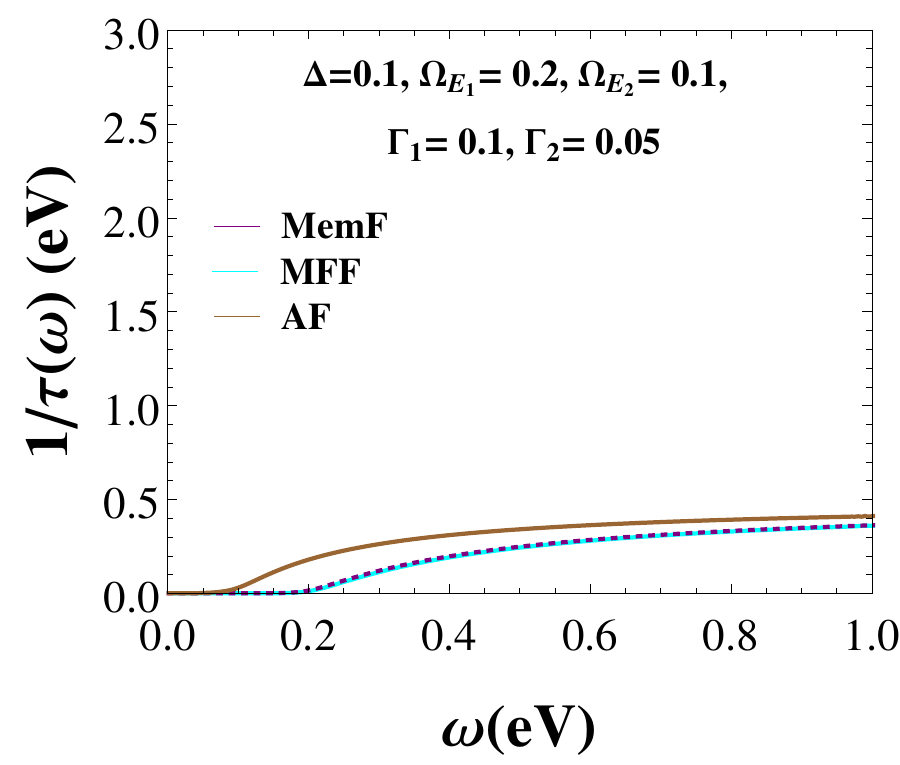}}
\subfigure[noonleline][]
{\label{fig:doubLoren12}\includegraphics[height=40mm,width=50mm]{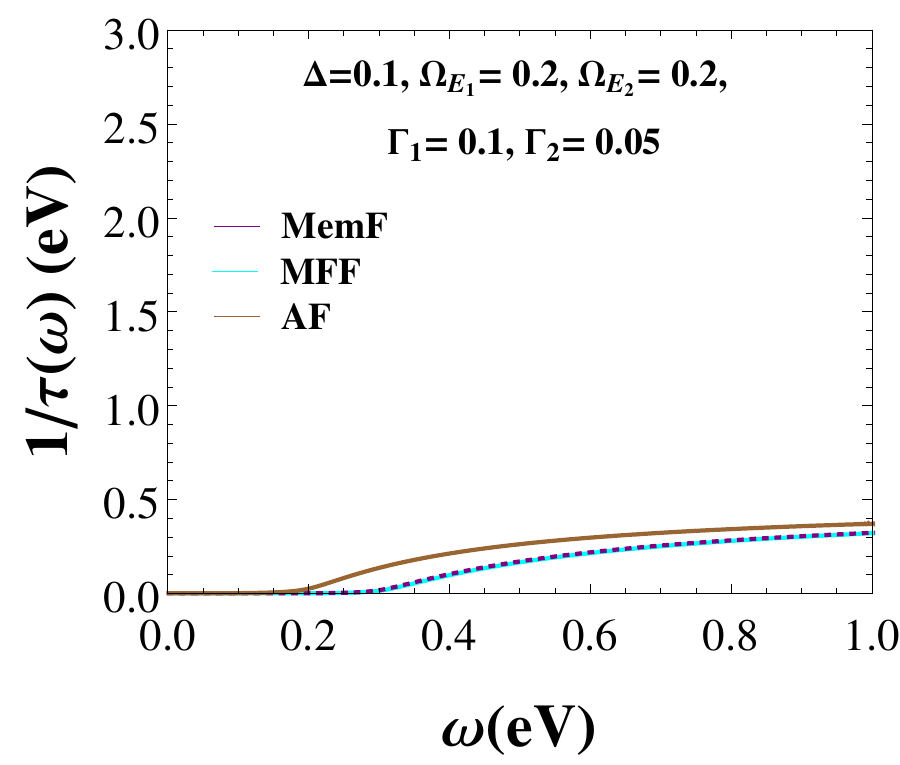}}
\caption{Scattering rate as a function of frequency with approaches namely Allen approach (AF; Brown), Mitrovi\'c-Fiorucci (MFF; Cyan), Memory (MemF; Purple) using Double Lorentzian function for phonon density of states. Here we consider the phonon peak frequencies $\Omega_{E_{1}}=0.2$ eV, $\Omega_{E_{2}}=0.05$, $0.1$ $0.2$ eV and Lorentzian widths $\Gamma_{1}=0.1$ eV, $\Gamma_{2}=0.01$, $0.05$ eV. For MFF and MemF, we have used gap value $\Delta=0.1$ eV.}
\label{fig:doubLorentB}
\end{figure*}
}

\newcommand{\figdoubLorentemp}{
\begin{figure*}
\centering
\hspace{0cm}
\subfigure[noonleline][]
{\label{fig:doubLorentemp1}\includegraphics[height=40mm,width=60mm]{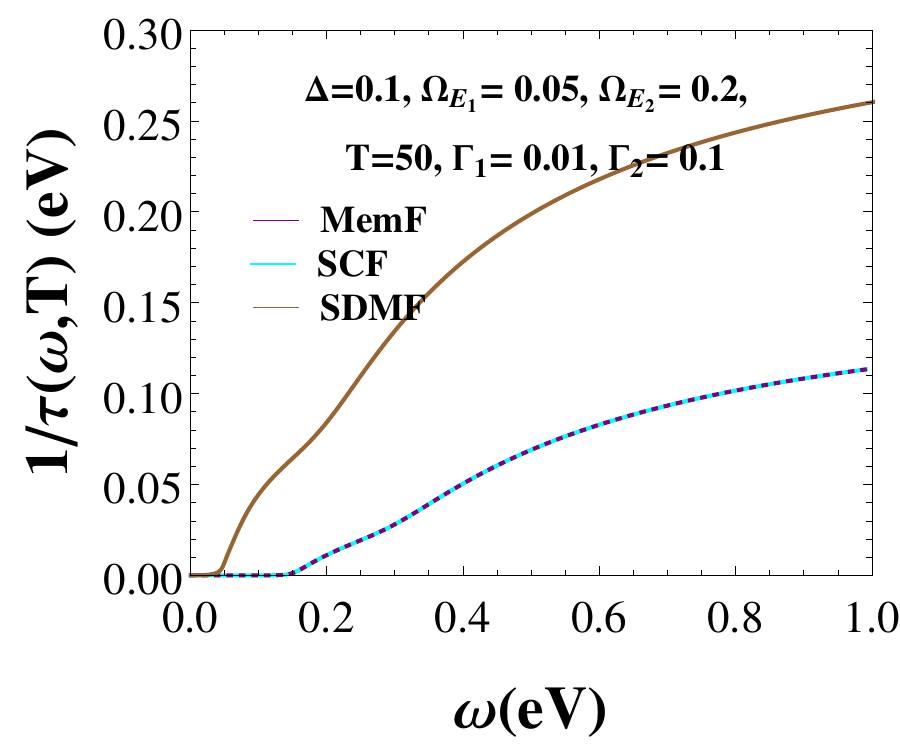}}
\hspace{0cm}
\subfigure[noonleline][]
{\label{fig:doubLorentemp2}\includegraphics[height=40mm,width=60mm]{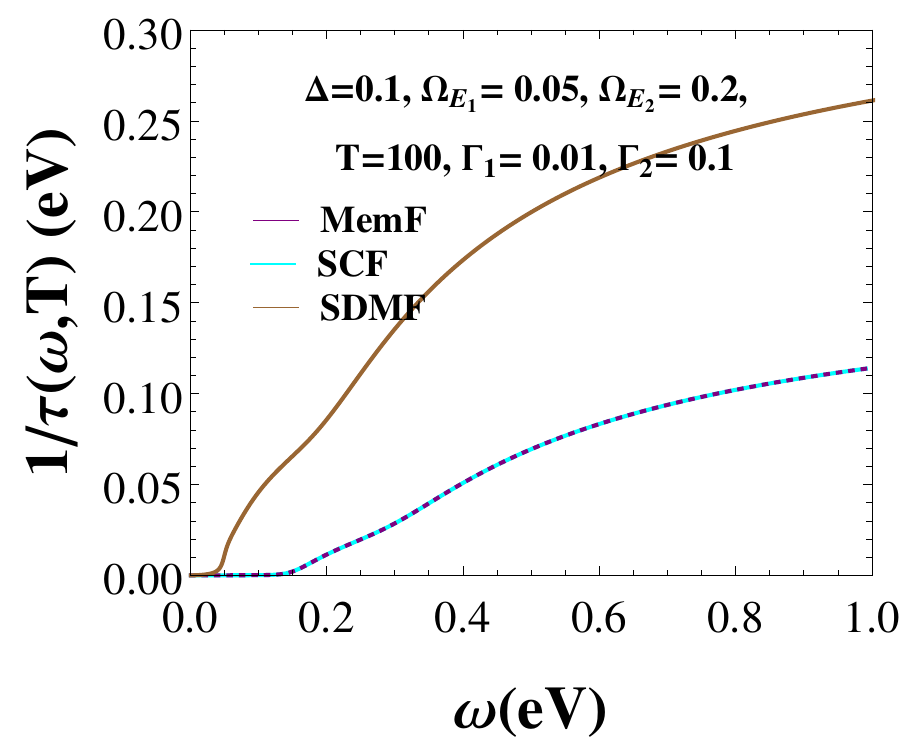}}
\subfigure[noonleline][]
{\label{fig:doubLorentemp3}\includegraphics[height=40mm,width=60mm]{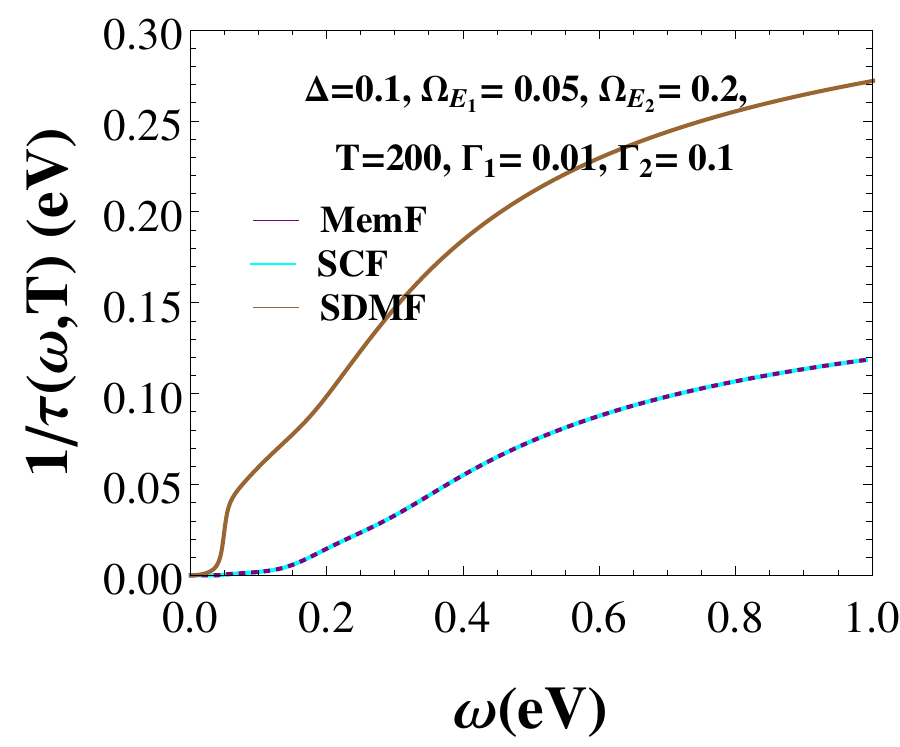}}
\subfigure[noonleline][]
{\label{fig:doubLorentemp4}\includegraphics[height=40mm,width=60mm]{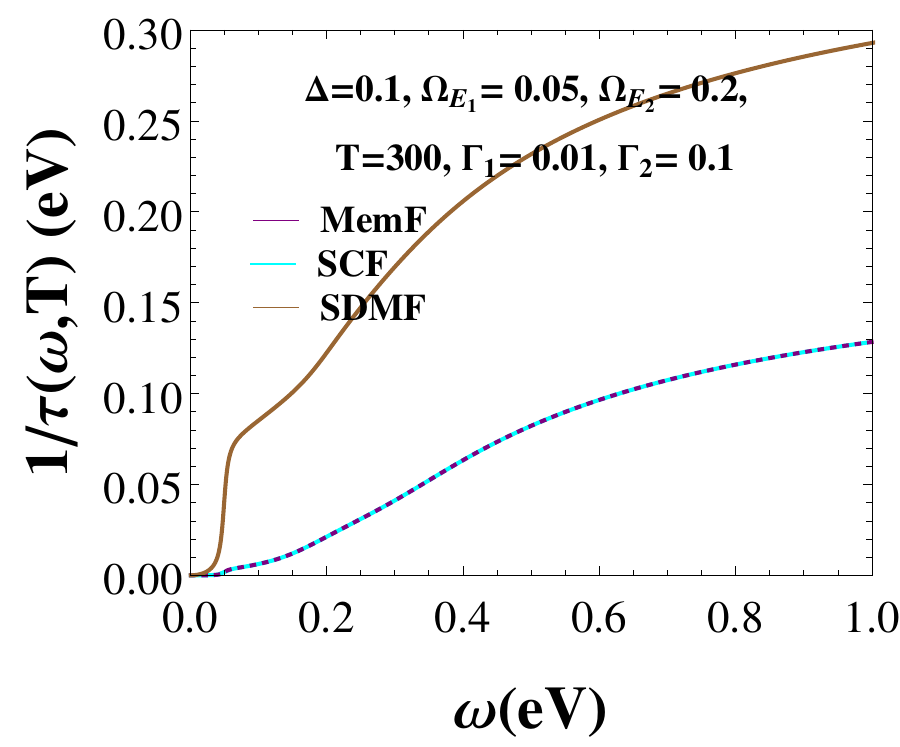}}
\caption{Temperature dependent scattering rate as a function of frequency with approaches namely Shulga et al. approach (SDMF; Brown), Sharapov-Carbotte (SCF; Cyan), Memory (MemF; Purple) using Double Lorentzian function for phonon density of states. Here we have fixed the phonon peak frequency $\Omega_{E_{1}}=0.05$, $\Omega_{E_{2}}=0.1$ eV, Lorentzian width $\Gamma_{1} = 0.01$, $\Gamma_{2} = 0.1$ eV and consider the gap gap value $\Delta=0.1$ eV for SCF and MemF. These plots are at different temperatures such as (a): $50$, (b): $100$, (c): $200$, (d): $300$ eV.}
\label{fig:doubLorentemp}
\end{figure*}
}

\begin{document}
\title{ A comparative study of finite frequency scattering rate from Allen, Mitrovi\'c-Fiorucci, Shulga-Dolgov-Maksimov, Sharapov-Carbotte and G\"otze-W\"olfle Memory Function formalisms}

\author{Pankaj Bhalla}
\email{pankajbhalla66@gmail.com}
\author{Navinder Singh}
\affiliation{Physical Research Laboratory, Navrangpura, Ahmedabad-380009 India.}

\begin{abstract}
We report a comparative study of the scattering rates using different formalisms such as Allen\cite{allen_71}, Shulga et al.\cite{shulga_91}, Mitrovi\'c et al.\cite{mitrovic_85}, Sharapov et al\cite{sharapov_05} and memory function formalism\cite{mori_65}. We discuss the frequency and the temperature dependent scattering rates for the case of electron-phonon interactions in these formalisms. The analysis has been done for different forms of phonon density of states (PDOS) and electron density of states (EDOS). An advantage of our study is that it shed light on the physical assumptions used in these formalisms.  From this detailed comparison, we observe that the memory function formalism is the most general one. All other formalisms are based on restrictive assumptions as discussed in this work.
\end{abstract}
\maketitle
\section{Introduction}
The study of the transport properties of strongly and weakly interacting systems such as metals, graphene, cuprates, etc. has attracted strong interest both in the theoretical and the experimental communities\cite{wilson_book, sarma_11, schrieffer_book, waldram_book, hussey_08, neto_09, navinder_book}. The behavior of these properties depend on the play of various interactions such as electron-impurity, electron-phonon, electron-electron interactions. Experimentally, these have been studied by several researchers using different probes such as dc resistivity, ac conductivity, magnetoresistance measurements, etc\cite{carbotte_11, timusk_99, basov_05, cooper_12, giannazzo_11, tan_07}. To understand these properties theoretically, several theories such as Drude, Boltzmann, etc\cite{wilson_book, ziman_book, basov_05, noelle_book} have been presented. Though these theories have explained several experimental facts with great success. However, as discussed in subsequent paragraphs, some of the properties are not addressed, thus questioning the basic assumptions on which these theories are based.

Drude's theory is the very first attempt to understand the mechanism of electrical conduction in metals. It successfully gives the expressions of electrical and thermal conductivities of metals and explains that the ratio of the electrical and thermal conductivity is directly proportional to the first power of temperature i.e. the Wiedemann-Franz law\cite{ashcroft_book}. But it fails to explain the temperature dependent behavior of electrical conductivity and thermal conductivity. Later, Bloch reformulates the Boltzmann theory where the electron-phonon interactions are considered and explains the temperature dependent behavior of the conductivity. The commonly used assumption in the solution of Bloch-Boltzmann equation is the relaxation time approximation (RTA)\cite{wilson_book, ziman_book, navinder_book}. The latter approximation means that the system will monotonically relax towards the local equilibrium state within a time scale called relaxation time. The general treatment of the Boltzmann equation is complicated and generally discussed numerically. In the traditional Bloch-Boltzmann equation, model used for phonon dispersion is the standard Debye model. Allen generalizes the Bloch-Boltzmann theory including the effects of general phonon density of states (PDOS), thus going beyond the Debye model\cite{allen_71}. It successfully explains the frequency dependent behavior of the electrical conductivity\cite{allen_71}. However, this formalism does not incorporate the temperature effects which are later introduced by Shulga, Dolgov and Maksimov (SDMF) within their formalism\cite{shulga_91}. Both these formalisms successfully reproduce the results predicted experimentally\cite{holstein_64, joyce_70, schlesinger_90}. But the above two formalisms (Allen and SDM) have not considered the concept of the non-constant electron density of states (EDOS). The latter concept is important to study the A15 systems\cite{mitrovic_85} and cuprates, graphene, etc\cite{bhalla_14, sarma_11}. This concept has been first addressed by Mitrovi\'c and Fiorucci\cite{mitrovic_85}. They study the effects of energy dependent electron density of states on several A15 compounds such as V$_{3}$Si, Nb$_{3}$Sn, V$_{3}$Ga\cite{mitrovic_85}. In a nutshell, this work is the generalization of the Allen formalism from constant EDOS to non-constant EDOS near the Fermi energy. It has been argued that this additional energy dependent behavior of the electron density of states yields the extra frequency dependence in the absorption spectra of A15 compounds. The formalism of Mitrovi\'c and Fiorucci is restricted to the zero temperature regime as is the Allen formalism. To study the temperature effects along with non-constant EDOS, Sharapov and Carbotte derive the expression of the scattering rate by considering both non-constant EDOS and at non-zero temperatures\cite{sharapov_05}. They derive the expression using a restrictive assumption which is presented in section \ref{sec:SCF}. To go beyond this assumption, an alternative formalism is needed. Such a formalism is introduced by Zwanzig and Mori in the context of time dependent correlation functions which are directly related to the transport properties\cite{zwanzig_61, zwanzig_book, mori_65}. This formalism is known as the memory function formalism\cite{forster_book, fulde_book}. Later, this is extended by G\"otze and W\"olfle to study the electrical conductivity of metals\cite{gotze_72}. It obtains the expression of the conductivity of metals that works well in all temperature and frequency regimes. The importance of this formalism is that it goes beyond the assumptions made in the previously discussed formalisms\cite{nabyendu_16a, bhalla_16aa}. In addition to these, this formalism has been widely used to study the electrical conductivity, thermal conductivity, Seebeck coefficient, Hall coefficient, etc. of various systems such as metals, graphene, cuprates\cite{lange_97, bhalla_16a, bhalla_16b, nabyendu_16b, bhalla_17a, bhalla_17b, lucas_15a, lucas_15b, patel_14, hartnoll_12, hartnoll_13, hartnoll_08, hartnoll_16, arfi_92, kupcic_17}.

In this context of transport properties, these different formalisms have been used several times with different perspectives. However, the comparative study of these formalisms at one place has not been addressed in the literature. The present paper remedies this situation. Here, we also perform the comparison using the different models for the phonon and electron density of states to analyze various formalisms and to check how restrictive assumptions effect the scattering rate.

This paper is organized as follows: First in Sec.~\ref{sec:formalisms}, we give the brief introduction of the different formalisms used to study the scattering rate. The brief picture of these formalisms such as Allen, Shulga-Dolgov-Maksimov, Mitrovic-Fiorucci, Sharapov-Carbotte and Memory function along with the sketch of their derivations and results has been presented. In Sec.~\ref{sec:comparison}, we perform the comparison of these formalisms based on the modeled density of states for electrons and phonons. Here we consider the Einstein and Lorentzian models for phonon density of states and square gap model for electron density of states. In Sec.~\ref{sec:analytical}, we derive the Allen and Mitrovi\'c-Fiorucci formalisms from the memory function formalism. Finally in Sec.~\ref{sec:conclusion}, we conclude.

\section{Formalisms}
\label{sec:formalisms}
In $1900$, Drude using ideas borrowed from the classical kinetic theory of gases formulated the expression for electrical conductivity\cite{drude_00} $\left(\sigma(\omega)=\sigma_{1}(\omega)+i\sigma_{2}(\omega)\right)$ as
\begin{equation}
\sigma(\omega) = \dfrac{\omega_{p}^2}{4\pi} \dfrac{1}{1/\tau - i\omega},
\end{equation}
where $1/\tau$ is the scattering rate of free electrons colliding with localized ion cores, $\omega_{p}^{2}= \frac{4\pi n e^2}{m}$ is the square of the plasma frequency, where $m$ is the electronic mass, $n$ is electron-density and $e$ is the electronic charge. For systems having electron-electron, electron-phonon interactions, this formula deviates due to the frequency dependent character of the scattering rate\cite{navinder_book, gotze_72}. Thus the formalism leads to the following generalization of the simple Drude formula:
\cite{gotze_72, timusk_03, puchkov_96}
\begin{equation}
\sigma(\omega, T) = \dfrac{\omega_{p}^2}{4\pi} \dfrac{1}{1/\tau(\omega, T) - i\omega (1+\lambda(\omega, T))},
\label{eqn:GDF1}
\end{equation}
called Generalized Drude Formula (GDF). Here $1/\tau(\omega, T)$ represents the frequency and the temperature dependent scattering rate and $\lambda(\omega, T)$ is the mass enhancement factor due to the electron-electron, electron-impurity and electron-phonon interactions. Generally, this expression is used to extract the relevant quantities like $1/\tau(\omega, T)$  and $\lambda(\omega, T)$ from the experimental data\cite{carbotte_11}. But to capture the behavior of the scattering rate analytically as predicted by experimental data, one has to derive a relation for the scattering rate and mass enhancement factor, starting from the microscopic Hamiltonian.

In the literature there are various formalisms that analytically derive $1/\tau(\omega, T)$ and $\lambda(\omega, T)$ from microscopic Hamiltonians. These formalisms are listed as follows: 
\begin{itemize}
\item[(A)]
Allen's formalism (AF)\cite{allen_71}
\item[(B)]
Shulga-Dolgov-Maksimov formalism (SDMF)\cite{shulga_91}
\item[(C)]
Mitrovi\'c -Fiorucci formalism (MFF)\cite{mitrovic_85}
\item[(D)]
Sharapov-Carbotte formalism (SCF)\cite{sharapov_05}
\item[(E)]
Memory function formalism (MemF)\cite{zwanzig_book, zwanzig_61, mori_65, gotze_72}
\end{itemize}
Brief overviews of these formalisms are discussed in the following subsections.
\subsection{The Allen Formalism (AF)}
\label{sec:AF}
\hspace*{-0.5cm}
\textbf{Introduction and Basic assumptions:}\\
In the case of metals, electrons can be treated as free electrons to a good approximation which undergo different collision mechanisms such as electron-phonon, electron-impurity, etc. These interactions govern the behavior of the transport properties. To deal with these, Allen gives the formalism to calculate the electrical conductivity by considering the lowest order effects of interactions on transport properties in local approximation\cite{allen_71}. \\
\textbf{Brief Derivation:}\\
The relation for optical conductivity has been derived using Fermi golden rule\cite{ashcroft_book}. Allen shows that for the collisionless case his theory yields the same result as given by semi-classical collisionless Boltzmann's approach\cite{allen_71}. To account collisions in his theory, he considers the following electron-phonon Hamiltonian 
\begin{equation}
H_{\textnormal{ep}}=\sum\limits_{\textbf{k}\textbf{k}'} M_{\textbf{k}\textbf{k}'}c^\dagger _{\textbf{k}'} c_\textbf{k}(a_\textbf{q}+a^\dagger_{-\textbf{q}}),
\label{eqn:AF1}
\end{equation}
where $M_{\textbf{k}\textbf{k}'}$ is the matrix element for the scattering of an electron from the Bloch state $\textbf{k}$ to $\textbf{k}'$, $b_\textbf{q}$ ($b_\textbf{q}^\dagger$) is an annihilation (creation) operator for a phonon of wave vector $\textbf{q}$ and corresponding energy $\omega_\textbf{q}$. With the above Hamiltonian and using {\it second order golden-rule formula}, the conductivity can be written as
\begin{eqnarray} \nonumber
\sigma(\omega)&=&\frac{2\pi e^2}{3\omega^3} \sum\limits_{\textbf{k}\textbf{k}'} \mid M_{\textbf{k}\textbf{k}'}\mid ^2 (\textbf{v}_{\textbf{k}}-\textbf{v}_{\textbf{k}^{\prime}})^2 f_{\textbf{k}}(1-f_{\textbf{k}^{\prime}})\\
&& \times \delta(\epsilon_{\textbf{k}^{\prime}} -\epsilon_{\textbf{k}} +\omega_{\textbf{q}} -\omega),
\label{eqn:AF2}
\end{eqnarray}
where $f_{\textbf{k}} = (e^{\beta\epsilon_{\textbf{k}}}+1)^{-1}$ is the Fermi distribution function and $\beta$ is the inverse of the temperature.\\
On performing these summations over $\textbf{k}$ and $\textbf{k}^{\prime}$ near to the Fermi surface and at $T=0$, the conductivity can be expressed as
\begin{equation}
\sigma(\omega)=\frac{\omega_p^2}{4\pi \omega}\left(\frac{1}{\omega \tau (\omega)}\right).
\label{eqn:AF3}
\end{equation}
Here $1/\tau(\omega)$, the frequency dependent scattering rate in the lowest order perturbation theory is given as
\begin{equation}
\frac{1}{\tau(\omega)}=\frac{2\pi}{\omega}\int_0^\omega d\Omega (\omega - \Omega)\alpha^{2}F(\Omega),
\label{eqn:AF4}
\end{equation}
where $\alpha^{2}F(\omega)$ is the phonon spectral function and is defined as
\begin{equation}
\alpha^{2}F(\omega)=\frac{N(0)}{4v_F^2}\langle \langle \mid M_{\textbf{k}\textbf{k}'}\mid ^2 (\textbf{v}-\textbf{v}')^2 \delta(\omega_{\textbf{q}} -\omega)\rangle\rangle.
\label{eqn:AF5}
\end{equation}
The above two equations are the central results of the Allen formulation. Here $N(0)$ is for the normalization, $v_{F}$ is the Fermi velocity and $\langle \langle \cdots \rangle\rangle$ represents the average over the constant energy shells corresponds to the momentum $\textbf{k}$ and $\textbf{k}'$.\\ 
\textbf{Results:}\\
This theory gives simple form for the frequency dependent scattering rate and gives good picture to explain the experimental data\cite{joyce_70, allen_71}. However, this theory has not explained the temperature dependent behavior of the scattering rate, thus the conductivity. This has been incorporated by SDMF\cite{shulga_91} which we have discussed in the next subsection.\\
\subsection{The Shulga-Dolgov-Maksimov Formalism (SDMF)}
\label{sec:SDMF}
\hspace*{-0.5cm}
\textbf{Introduction and Basic assumptions:}\\
In their formalism, Shulga et al. consider metallic system as in AF. They argue that strong electron-phonon interactions can lead to non-Fermi liquid properties. Thus they attempt to address electronic system in high temperature cuprate superconductors and their non-Fermi liquid properties in the normal states\cite{shulga_91}. The SDMF in essence extends Allen's treatment to include finite temperature effects (recall that AF is for $T=0$K).\\
\textbf{Brief Derivation:}\\
In this context, the frequency dependent optical conductivity for the case of the electron-phonon interaction is calculated based on the Green's function approach\cite{mahan_book}. They start with the Kubo formula of optical conductivity
\begin{eqnarray}
 \sigma_{\mu\nu}(\omega) &=& \frac{K_{\mu\nu}(\omega)}{4\pi i \omega},
 \label{eqn:SMa}
\end{eqnarray}
where 
\begin{eqnarray}\nonumber
 K_{\mu\nu}(\omega) &=& -4\pi i e^{2} \int \frac{d^{4}k}{(2\pi)^{4}} \gamma_{\mu}(k,k+q) G(k+q) \\
 && \times \Gamma_{\nu}(k,k+q)G(k).
 \label{eqn:SMb}
\end{eqnarray}
In the above equation, $G(k)$ is the single electron Green's function, $\gamma_{\mu}$ is the bare vertex, $\Gamma_{\nu}$ is the dressed vertex and ($\mu, \nu$) represents the Cartesian coordinates. With lengthy calculation, they show that in an isotropic system the optical conductivity is expressed as
\begin{eqnarray} \nonumber
 \sigma(\omega, T) &=& \frac{\omega_{p}^{2}}{4\pi i \omega} \bigg\lbrace \int_{-\omega}^{0} d\omega' \tanh \bigg(\frac{\omega + \omega'}{2T} \bigg) S^{-1}(\omega, \omega')\\ \nonumber
 && + \int_{0}^{\infty} d\omega' \bigg[\tanh\bigg(\frac{\omega + \omega'}{2T} \bigg) - \tanh\bigg(\frac{\omega'}{2T} \bigg)  \bigg]\\
 && \times S^{-1}(\omega, \omega') \bigg\rbrace,
 \label{eqn:SM1}
\end{eqnarray}
where $S(\omega, \omega')$ is defined as
\begin{eqnarray}
 S(\omega, \omega') &=& \omega + \Sigma^{*}(\omega + \omega') - \Sigma(\omega').
 \label{eqn:SM2}
\end{eqnarray}
Here $\Sigma(\omega) (=\Sigma_{1}(\omega) + i \Sigma_{2}(\omega))$ is the single particle self energy and $T$ is the temperature. The real and imaginary parts of the self energy are defined as\cite{vonsovosky_book}
\begin{eqnarray}\nonumber
 Re[\Sigma(\omega, T)] &=& - \int d\omega' \alpha^{2}F(\omega') Re\bigg[\psi\bigg(\frac{1}{2} + i \frac{\omega+\omega'}{2\pi T} \bigg)\\
 && - \psi\bigg(\frac{1}{2} + i \frac{\omega-\omega'}{2\pi T} \bigg)  \bigg].
 \label{eqn:SM3}
\end{eqnarray}
\begin{eqnarray}\nonumber
 Im[\Sigma(\omega, T)] &=& - \frac{\pi}{2} \int d\omega' \alpha^{2}F(\omega') \bigg[ 2\coth\bigg(\frac{2\omega'}{2T}  \bigg)\\ \nonumber
 && - \tanh\bigg(\frac{\omega+\omega'}{2T}  \bigg) + \tanh\bigg(\frac{\omega-\omega'}{2T}  \bigg)  \bigg],\\
 \label{eqn:SM4}
\end{eqnarray}
with $\alpha^{2}F(\omega)$ as the electron-phonon spectral function and $\psi(\omega)$ is the logarithmic derivative of the gamma function.\\
Substituting Eqs.~(\ref{eqn:SM3}), (\ref{eqn:SM4}) in Eq.~(\ref{eqn:SM2}) and then comparing with the Generalized Drude formula Eq.~(\ref{eqn:GDF1}), the frequency and the temperature dependent scattering rate can be expressed as\cite{shulga_91}
\begin{eqnarray}\nonumber
 \frac{1}{\tau(\omega,T)} &=& \frac{\pi}{\omega} \int d\omega' \alpha^{2}F(\omega') \bigg\lbrace 2\omega \coth\bigg(\frac{\omega'}{2T}\bigg) \\ \nonumber
 && - (\omega + \omega') \coth\bigg(\frac{\omega + \omega'}{2T}\bigg)\\
 && + (\omega - \omega') \coth\bigg(\frac{\omega - \omega'}{2T}\bigg)\bigg\rbrace.
 \label{eqn:SM5}
\end{eqnarray}
Taking temperature $T=0$ in the above expression, we come back to the expression Eq.~(\ref{eqn:AF5}) given by Allen, as it should\cite{allen_71}.\\
\textbf{Results:}\\
This formalism explains the finite temperature behavior of the scattering rate. It also explains that at high temperatures and at characteristic phonon frequency (Debye frequency) it is not possible to describe the cuprate systems using standard Fermi liquid theory. It successfully calculate the conductivity expression using the many body approach and gives agreement of the experimentally measured reflectivity data in the far infrared region\cite{schlesinger_90}.

In both the above discussed formalisms, the expressions for the scattering rate have been derived taking only the phonon spectral function. However, the electron density of states is not incorporated in them. It is assumed constant near the Fermi energy. This is an important aspect to explain the transport properties of compounds such as A15 superconductors, cuprates\cite{mitrovic_83, kieselmann_82, wolf_80}. Thus these formalisms require further generalization. To account these, the first step has been taken by Mitrovi\'c and Fiorucci in their formalism which we discuss in the next subsection.
\subsection{The Mitrovi\'c - Fiorucci Formalism (MFF)}
\label{sec:MFF}
\hspace*{-0.5cm}
\textbf{Introduction and Basic assumptions:}\\
For systems such as  V$_{3}$Si, Nb$_{3}$Sn, V$_{3}$Ga, there is a rapidly varying electronic density of states (EDOS) near the Fermi energy which effects the experimental measurement of the coupling function $\alpha^2F(\omega)$, because it is deduced by numerical inversion technique\cite{mcmillan_65}. This requires the detailed calculation of the scattering rate taking non-constant EDOS in the vicinity of the Fermi level. Mitrovi\'c and Fiorucci consider this problem and derive the expression of the scattering rate\cite{mitrovic_85}. They have also considered the same assumptions as by Allen, but with one important point i.e. the non-constant EDOS at the Fermi energy.\\
\textbf{Brief Derivation:}\\
In calculation, they consider the expression of optical conductivity as derived by Allen in Eq.~(\ref{eqn:AF2}). Then introducing energy integrals in the expression and doing algebraic manipulations, it has been found that the optical conductivity comes same as given in Eq.~(\ref{eqn:AF3}). Here the difference is in the frequency dependent scattering rate which includes the energy dependent electronic density of states. It is expressed as\cite{mitrovic_85}
\begin{equation}
\frac{1}{\tau(\omega)} = \frac{2\pi}{\omega} \int_{0}^{\omega} d\Omega  \alpha^{2}F(\Omega) \int_{0}^{\omega-\Omega} d\epsilon \frac{1}{2} \left[\frac{N(- \epsilon)}{N(0)} + \frac{N(\epsilon)}{N(0)} \right],
\label{eqn:MF1}
\end{equation}
where $N(\epsilon)$ is the energy dependent electronic density of states and $\alpha^{2}F(\Omega)$ is same as Eq.~(\ref{eqn:AF5}). If we consider $N(\epsilon)$ as a constant, it yields the same expression of scattering rate as given by Eq.~(\ref{eqn:AF4}).\\
\textbf{Results:}\\
This formula explains that the extra energy dependence in experimental data for the electron-phonon coupling system comes from $N(\epsilon)$ i.e. the energy dependent behavior of electronic density of states\cite{mcknight_79, mitrovic_84}. However, this approach is valid only at zero temperature case. Thus further generalization was needed and it is given by Sharapov-Carbotte formalism\cite{sharapov_05}.
\subsection{The Sharapov-Carbotte Formalism (SCF)}
\label{sec:SCF}
\hspace*{-0.5cm}
\textbf{Introduction and Basic assumptions:}\\
To study the temperature dependent behavior of the ac conductivity, Sharapov and Carbotte develop an approach in which the conductivity is expressed through quasiparticle self energy. This formalism is based on the following assumptions\cite{sharapov_05}. One assumption is the independence of self energy from quasiparticle momentum (only frequency dependence is considered). The other assumption is that the magnitude of the electron self energy difference at two different frequencies is smaller than the frequency i.e. $\vert \Sigma(\epsilon+\omega) - \Sigma(\epsilon) \ll \omega \vert$\cite{sharapov_05}.\\
\textbf{Brief Derivation:}\\
Under these assumptions, they derive the expression for temperature and frequency dependent scattering rate. The calculation starts with the Kubo formula\cite{kubo_57} and the current-current correlation function is evaluated using Green function technique\cite{mahan_book}. In the calculation, the two particle Green's function is written as a product of the two dressed single particle Green's functions. This obtains the real part of optical conductivity as\cite{lee_89, kaufmann_98}
\begin{eqnarray} \nonumber
\textnormal{Re} \left[ \sigma(\omega,T) \right] &=& \frac{\omega_{p}^{2}}{4 \pi} i \textnormal{Im} \bigg[  \int_{-\infty}^{\infty} d\omega^{\prime} \bigg(\frac{f(\omega^{\prime})-f(\omega+\omega^{\prime})}{\omega} \bigg) \\
&& \frac{1}{\omega-\Sigma(\omega)+\Sigma^{*}(\omega^{\prime})} \bigg].
\label{eqn:SC1}
\end{eqnarray}
Here $f(\omega)$ is the Fermi distribution function and $\Sigma(\omega)$ is the self energy due to electron-phonon/Boson interactions.\\
The imaginary part of the self energy due to electron-phonon/Boson interaction is\cite{sharapov_05}
\begin{eqnarray} \nonumber
\textnormal{Im} \Sigma(\omega,T) &=& - \pi \int_{0}^{\infty} d\Omega \alpha^{2}F(\Omega)\\ \nonumber
&& \times \bigg[ \frac{\tilde{N}(\omega-\Omega)}{N(0)}\big(n(\Omega)+1-f(\omega-\Omega)\big) \\
&& + \frac{\tilde{N}(\omega+\Omega)}{N(0)}\big(n(\Omega)+f(\omega+\Omega)\big) \bigg],
\label{eqn:SC2}
\end{eqnarray}
where $n(\omega) = \bigg(e^{\omega/T} -1 \bigg)^{-1}$ is the phonon/Boson distribution function and $N(\omega)$ is the quasiparticle density of states.\\
The above self energy expression is substituted in the optical conductivity expression (Eq.~(\ref{eqn:SC1})). Then, the optical conductivity is written in the form of Generalized Drude formula (Eq.~\ref{eqn:GDF1}). From the resulting expression, the generalized Drude scattering rate is obtained which can be expressed as\cite{sharapov_05}
\begin{eqnarray} \nonumber
\frac{1}{\tau(\omega,T)} &=& \frac{\pi}{\omega} \int_{0}^{\infty} d\Omega \alpha^{2}F(\Omega) \int_{-\infty}^{\infty} d\omega^{\prime} \\\nonumber
&& \times \bigg[ \frac{\tilde{N}(\omega^{\prime}-\Omega)}{N(0)} + \frac{\tilde{N}(-\omega^{\prime}+\Omega)}{N(0)} \bigg] \\\nonumber
&& \times \big[n(\Omega)+ f(\Omega-\omega^{\prime}) \big] \big[f(\omega^{\prime}+\omega)-f(\omega^{\prime}+\omega) \big].\\
\label{eqn:SC3}
\end{eqnarray}
This is the final expression of the frequency and temperature dependent scattering rate which contains both phonon/boson and electron density of states. Here $\alpha^{2}F(\omega)$ has same meaning as expressed by Eq.~(\ref{eqn:AF5}).\\ 
\textbf{Results:}\\
This formula has been widely used by authors to extract $\alpha^{2}F(\omega)$ from the experimental data\cite{hwang_11, hwang_12}. The results from these extractions of $\alpha^{2}F(\omega)$ can not be fully trusted as the formula (Eq.~\ref{eqn:SC3}) itself is based on several restrictive assumptions as mentioned, especially in the low frequency regime\cite{bhalla_16aa}. To go beyond these assumptions, we discuss memory function technique for generalized scattering rate in the next subsection.
\subsection{Memory function Formalism}
\label{sec:MemF}
\hspace*{-0.65cm}
\textbf{Introduction and Basic assumptions:}\\
The memory function formalism is introduced by Zwanzig and Mori to study the time dependent correlation functions which are directly connected to the transport properties\cite{zwanzig_book, zwanzig_61, mori_65}. By using projection operators\cite{fulde_book, forster_book}, the Kubo formula for the generalized susceptibility or AC electrical conductivity in the present case can be written as 
\begin{eqnarray}\nonumber
 \sigma(z, T) &=& i\frac{\omega_{p}^{2}}{4\pi} \frac{1}{z+M(z,T)}; \hspace*{0.2cm} M(z,T) = \frac{z\chi(z)}{\chi_{0}-\chi(z)},\\
 \label{eqn:Mem1}
\end{eqnarray}
which is an exact expression within the linear response theory. Here $z = \omega + i\eta$($\eta \rightarrow 0^{+}$) is the complex frequency and  $M(z, T)$ is the memory function. The latter has two parts where the imaginary part is equivalent to the scattering rate and the real part contribute to the mass enhancement factor\cite{fulde_book}. The other symbol $\chi(z)$ is the current-current correlation function and $\chi_{0}$ is the static limit of the former correlation function. Here the exact computation of $M(z,T)$ starting from a microscopic Hamiltonian is not possible. However, in normal circumstances electron-phonon interaction can be treated as a perturbation. A perturbative expansion of the memory function is developed by G\"otze and W\"olfle which is based on this assumption as explained below.\\
\textbf{Brief Derivation:}\\
Based on the above mentioned description, G\"otze and W\"olfle derive the expression of the electrical conductivity using the total Hamiltonian
\begin{eqnarray}
 H &=& H_{e} + H_{ep} + H_{p}.
\end{eqnarray}
Here $H_{e} = \sum_{\textbf{k},\sigma} \epsilon_{\textbf{k}} c_{\textbf{k}\sigma}^{\dagger} c_{\textbf{k}\sigma}$, $H_{p} = \sum_{\textbf{q}} \big( \omega_{q} + \frac{1}{2}\big) b_{\textbf{q}}^{\dagger}b_{\textbf{q}}$ and $H_{ep}$ is defined in Eq.~(\ref{eqn:AF1}), $\omega_{q}$ is the phonon frequency, $\epsilon_{\textbf{k}}$ is the electron dispersion and $b_{q}^{\dagger}(b_{q})$ is the phonon creation (annihilation) operator. They start with the definition of correlation function
\begin{eqnarray}
 \chi(z,T) &=& \langle \langle J;J \rangle\rangle_{z} = -i \int_{0}^{\infty} dt e^{izt} \langle [J(t);J]\rangle.
\end{eqnarray}
Here $J = m^{-1} \sum_{k} e\textbf{k}\cdot \hat{n} c_{\textbf{k}\sigma}^{\dagger} c_{\textbf{k}\sigma}$ is the electrical current, $[ , ]$ is the commutator between two quantities, $\langle\langle \cdots \rangle\rangle $ represents the ensemble average and the Fourier transform. Then, on using the equation of motion and algebraic calculations, they find that the memory function becomes
\begin{eqnarray} \nonumber
 M(z,T) &=& \frac{m}{N_{e}}\frac{\langle \langle [J,H]; [J,H]\rangle\rangle_{z=0} - \langle \langle [J,H]; [J,H]\rangle\rangle_{z}}{z}.\\
 \label{eqn:Mem2}
\end{eqnarray}
Substituting the total Hamiltonian in the above equation and then solving, $M(z,T)$ becomes
\begin{eqnarray} \nonumber
 M(z,T) &=& \frac{2}{3} \frac{1}{m N_{e}} \sum_{\textbf{k}, \textbf{k}'}\vert D(\textbf{k}-\textbf{k}') \vert ^{2}(\textbf{k}-\textbf{k}')^{2} \\ \nonumber
 && (f(1-f')(1+n) - f'(1-f)n) 
\\\nonumber && \times \frac{1}{\epsilon_{\textbf{k}}-\epsilon_{\textbf{k}'}-\omega_{\textbf{k}-\textbf{k}'}} \bigg[ \frac{1}{(\epsilon_{\textbf{k}}-\epsilon_{\textbf{k}'}-\omega_{\textbf{k}-\textbf{k}'}+ z)}\\
&& - \frac{1}{(\epsilon_{\textbf{k}}-\epsilon_{\textbf{k}'}-\omega_{\textbf{k}-\textbf{k}'}- z)} \bigg].
\end{eqnarray}
On performing analytic continuation, the imaginary part of the memory function or the scattering rate is given by
\begin{eqnarray}\nonumber
 Im[M(z,T)] &=&  \frac{1}{\tau(\omega, T)} \\ \nonumber
 &=& \frac{2\pi}{\omega} \int_{0}^{\infty} d\omega' \alpha^{2}F(\omega') \int_{-\infty}^{\infty} d\epsilon  N(\epsilon) \\ \nonumber
 && \times \big[1-f(\epsilon-\epsilon_{F})\big] n(\omega') \\ \nonumber
&& \times  \big[ N(\epsilon-\omega'+\omega) f(\epsilon-\epsilon_{F}-\omega'+\omega) \\ \nonumber
&& \times \big[e^{\beta \omega}-1\big] - (\text{terms with $\omega \rightarrow -\omega$}) \big].\\
\label{eqn:Mem3}
\end{eqnarray}
This is the general expression of the frequency and the temperature dependent scattering rate. \\
\textbf{Results:}\\
The formula derived within this formalism has successfully reproduced several results which are in agreement with the experimental findings\cite{wilson_book, gotze_72}. In addition to this, it has also been used to study both the gapped and non-gapped systems\cite{bhalla_16aa}. This formalism is applicable to study the scattering rate at all temperature and frequency regimes as the earlier formalisms has limited range to study the scattering rate. Thus, this formalism is a good choice to study the transport properties of various systems such as metals, graphene, cuprates, etc\cite{navinder_book, nabyendu_16a, nabyendu_16b, bhalla_16b, bhalla_17a, bhalla_17b}. 
\onecolumngrid
\figDel
\twocolumngrid
For a detailed analysis of these formalisms, we discuss  in the next section the comparison between them based on the modeled density of states for the electrons and phonons.

\section{Comparison of scattering rates based on modeled density of states}
\label{sec:comparison}
To compare the formalisms, we plot the scattering rates, expressed in Eqs.~(\ref{eqn:AF4}), (\ref{eqn:SM5}), (\ref{eqn:MF1}), (\ref{eqn:SC3}) and (\ref{eqn:Mem3}) using different forms of $\alpha^{2}F(\omega)$ and $N(\omega)$ at different parameter values such as temperature, frequency, etc. This analysis is presented in two cases. One corresponds to the zero temperature regime and another corresponds to the finite temperature regime. In each case, we present two subcases corresponding to different PDOS.
\subsection{Zero temperature study (comparing AF, MFF and MemF formalisms)}
In this subsection, we discuss the scattering rates at zero temperature using the approaches such as AF, MFF and MemF formalisms. The scattering rates are computed with different models for PDOS namely Einstein and Lorentzian model.
\subsubsection{Einstein/Delta model for PDOS}
Here we consider the Einstein model for $\alpha^{2}F(\Omega)$ which is given as
\begin{eqnarray}
\alpha^{2}F(\Omega) &=& \Omega_{E}\delta(\Omega - \Omega_{E}),
\label{eqn:EM}
\end{eqnarray}
where $\Omega_{E}$ is the phonon peak frequency.

Substituting the above expression in Eqs.~(\ref{eqn:AF4}), (\ref{eqn:MF1}) and (\ref{eqn:Mem3}), the scattering rate is computed and is shown in Fig.~\ref{fig:Del}. Here for the MFF and the MemF formalisms, we consider the square well type model for the electron density of states. According to this model, the density of states is zero within the regime $-\Delta$ to $\Delta$ and unity outside this regime. The motivation to consider this simplified model is that here we are mainly concerned about the behavior of the scattering rate using constant and non-constant electron density of states (EDOS). In Fig.~\ref{fig:Del}, the frequency dependent scattering rate is shown at fixed gap value $\Delta = 0.1$ eV as a function of frequency $\omega$. Here we have presented $1/\tau(\omega)$ at different phonon peak frequency $\Omega_{E} = 0.05$, $0.1$ and $0.2$ eV in Figs.~\ref{fig:Del1}, \ref{fig:Del2} and \ref{fig:Del3} respectively and the scattering rate by AF, MFF and MemF is represented as brown, cyan and purple respectively. We observe three features. One is the vanishing scattering rate below the phonon peak frequency due to the zero PDOS. Second is the effect of the gap value in $1/\tau(\omega)$ computed using the MFF and MemF formalisms. Due to this gap scenario, the scattering rate remains zero not only up to the value of $\Omega_{E}$, but to $\Omega_{E} + \Delta$ as it should. Third is the increasing magnitude of the scattering rate with the increase in the phonon peak frequency due to the participation of the high frequency phonons. In addition to these, we also find that MemF and the MFF formalisms agree with each other very well. However, the AF show deviation from these two due to the lack of the effects of electron density of states, as it considers a constant EDOS.
\subsubsection{Lorentzian model for PDOS}

In this case, we consider that the phonon density of states is of the form of Lorentzian function and is expressed as
\begin{eqnarray}
 \alpha^{2}F(\Omega) &=& \frac{\Omega^{2}\Omega_{p}^{2}}{\big(\Omega^{2} - \Omega_{E}^{2}\big)^{2}+\big(\Gamma \Omega\big)^{2}}.
 \label{eqn:LM}
\end{eqnarray}
\onecolumngrid
\figLorent 
\twocolumngrid
Here $\Omega_{p}$ is the plasma frequency, $\Omega_{E}$ is the Lorentzian peak frequency and $\Gamma$ is the Lorentzian width. Using this and square well type model for EDOS in Eqs.~(\ref{eqn:AF4}), (\ref{eqn:MF1}) and (\ref{eqn:Mem3}), we have plotted the scattering rate as a function of frequency at fixed gap value $\Delta = 0.1$ eV and $\Omega_{p} = 0.05$ eV. Here the plots are presented at different $\Omega_{E}$ such as $0.05$, $0.1$ and $0.2$ eV and $\Gamma$ such as $0.01$ and $0.05$ eV. It is observed that the scattering rate remains negligible up to the certain value of the frequency depending on the value of the Lorentzian peak. Then it shows sudden increase in the scattering rate due to the phonon spectral function. At the high frequencies, it approaches to the saturation behavior. We also find that the increase in the Lorentzian peak and the Lorentzian width, there is a decrease in the scattering rate as shown in Fig.~\ref{fig:Lorent}.
\subsection{Finite temperature study (comparing SDMF, SCF, MemF formalisms)}
Here we discuss the scattering rates at finite temperatures using SDMF, SCF and MemF formalisms. Similar to the previous case, we consider the two models for PDOS. The AF and MFF are manifestly zero temperature formalisms.
\subsubsection{Einstein/Delta model for PDOS}
\onecolumngrid
\figDeltemp
\figLorentemp
\twocolumngrid
Considering the Einstein/Delta function model for PDOS Eq.~(\ref{eqn:EM}),  the scattering rate is shown at different temperatures such as $T = 50$, $100$, $200$ and $300$ K in Fig.~\ref{fig:Deltemp}. Here $1/\tau(\omega,T)$ is plotted using SDMF, SCF and MemF formalisms. The electronic density of states has been considered same as in the previous case and the phonon peak frequency $\Omega_{E} = 0.05$ eV and $\Delta = 0.1$ eV are kept fixed. It is observed that the magnitude of the scattering rate increases with the rise in temperature. Also, at high temperatures and small frequency, the scattering rate show some finite value and overcomes the signature of the gap. This is due to electrons excited above the energy gap. Looking at Fig.~\ref{fig:Deltemp1}, the scattering rate around $0.1$ eV is negligibly small. The reason is the small thermal energy $0.004$ eV in comparison to the phonon energy $0.05$ eV. This results the lack of phonon excitations to give finite scattering rate. In contrary to this in Fig.~\ref{fig:Deltemp4}, with the thermal energy $0.03$ eV which is near to the phonon energy there is higher scattering rate due to sufficient phonon excitations. This whole scenario is clearly visible within the SDMF approach where no gap scenario is considered. In the case of MemF and SCF which include the gap, the gap  signature is clearly visible in Fig.~\ref{fig:Deltemp1}. In fig.~\ref{fig:Deltemp4}, due to the less thermal energy in comparison to the excitation energy required for electrons and phonons excitations, the scattering rate does not show the sufficient finite value as by SDMF approach (shown in brown color). Also the gap signature is not clearly visible due to the energy comparisons as discussed earlier.
\subsubsection{Lorentzian model for PDOS}

Similarly for the Lorentzian model for PDOS Eq.~(\ref{eqn:LM}), the temperature dependent behavior of the scattering rate is shown at $T= 50$, $100$, $200$ and $300$ K and at fixed $\Omega_{E} = 0.05$, $\Gamma = 0.01$ and $\Delta = 0.1$ eV in Fig.~\ref{fig:Lorentemp}. Here we observe the same features related to the temperature as discussed earlier in the case of Einstein model. A more realistic, but little bit cumbersome case of double well PDOS is discussed in Appendix~\ref{app:1} 

\section{Derivation of AF and MFF from MemF formalism}
\label{sec:analytical}
In addition to this comparison, we also perform the consistency check of the scattering rate calculated by the memory function with other approaches as below. For this we have considered the simplest model square well type for the EDOS in our calculation process and perform calculations only for the temperature independent formalisms.\\
According to the memory function approach, it is calculated that the scattering rate using the square type model for the electron density of states is expressed as
\begin{eqnarray}\nonumber
\frac{1}{\tau(\omega, T)} &=& \frac{\pi^{3} N^{2} \rho_{F}^{2}}{4mk_{F}^{5}} \int_{0}^{q_{D}} dq q^{3} \vert D(q) \vert ^{2}\frac{1}{\beta}\\ \nonumber
&& \times \bigg[\frac{1}{e^{\beta\omega_{q}}-1} \bigg( \frac{e^{\beta\omega}-1}{\omega} \bigg) \bigg(\frac{1}{e^{\beta(\omega-\omega_{q})}-1} \bigg)\\ \nonumber
&& \log \bigg(\frac{1+e^{-\beta (-\omega + \omega_{q})}}{1+e^{\beta(\omega_{q} - \omega)}} \bigg) + (\textnormal{terms with} \, \omega \rightarrow -\omega)\bigg].\\
\label{eqn:Comp1}
\end{eqnarray}
The above expression in different limits of gap and temperature is discussed as follows:\\
\textbf{Case-I (With Allen's approach)} \\
On substituting $\Delta = 0$ in the above equation and simplifying, the scattering rate becomes
\begin{eqnarray}\nonumber
\frac{1}{\tau(\omega, T)} &=& \frac{\pi^{3} N^{2} \rho_{F}^{2}}{4mk_{F}^{5}} \int_{0}^{q_{D}} dq q^{3} \vert D(q) \vert ^{2} \bigg[\frac{(\omega-\omega_{q})}{\omega} \frac{1}{e^{\beta\omega_{q}}-1}\\ \nonumber
&& \times \bigg( \frac{e^{\beta\omega}-1}{e^{\beta(\omega-\omega_{q})}-1} \bigg) + (\textnormal{terms with} \, \omega \rightarrow -\omega)\bigg].\\
\end{eqnarray}
Now on taking the limit $T \rightarrow  0$ i.e. $\beta \rightarrow \infty$, the above equation becomes
\begin{equation}
\frac{1}{\tau(\omega)} = \frac{\pi^{3} N^{2} \rho_{F}^{2}}{4mk_{F}^{5}}  \int_{0}^{q_{D}} dq q^{3} \vert D(q) \vert ^{2}\frac{(\omega-\omega_{q})}{\omega}.
\end{equation}
Further on substitute $q=\omega_{q}/c_{s}$ in above equation and changing the variable $\omega_{q}$ to $\Omega$, we have 
\begin{equation}
\frac{1}{\tau(\omega)} = \frac{2\pi}{\omega}  \int_{0}^{\omega_{D}} d\Omega \alpha^{2}F(\Omega) (\omega-\Omega).
\end{equation}
Here $\alpha^{2}F(\Omega)$ is the phonon spectral function which is defined as
\begin{equation}
\alpha^{2}F(\Omega) = \frac{\pi^{2} N^{2} \rho_{F}^{2}}{8mk_{F}^{5}c_{s}^{4}}\Omega^{3} \vert D(\Omega) \vert ^{2}.
\end{equation}
Also the upper cutoff limit in the integration is $\omega_{D}$. But for phonon excitations $\omega \geq \omega_{D}$. Hence, the maximum value of upper cutoff will be replaced by $\omega$. Hence with this argument, the above equation becomes 
\begin{equation}
\frac{1}{\tau(\omega)} = \frac{2\pi}{\omega}  \int_{0}^{\omega} d\Omega \alpha^{2}F(\Omega) (\omega-\Omega).
\label{memallen}
\end{equation}
Thus the imaginary part of memory function (also called as scattering rate) comes similar to the Allen's formula for scattering rate Eq.~(\ref{eqn:AF4}) for electron-phonon interaction in the case of metals.\\
\textbf{Case-II (With Mitrovi\'c and Fiorucci approach)} \\
To discuss our case with $N(\epsilon) \neq$ constant and zero temperature, take $T \rightarrow 0$ in Eq.~(\ref{eqn:Comp1}), we get
\begin{eqnarray}\nonumber
\frac{1}{\tau(\omega, T)} &=& \lim_{\beta\to\infty} \frac{\pi m}{3N_{e}} \int_{0}^{q_{D}} dq q^{3} \vert D(q) \vert ^{2}\frac{1}{\beta}\bigg[n  \frac{e^{\beta\omega}-1}{\omega} \\ \nonumber
&&  \times \frac{1}{e^{\beta(\omega-\omega_{q})}-1} \log \bigg(\frac{1+e^{-\beta(\Delta-\omega+\omega_{q})}}{1+e^{-\beta(\Delta+\omega-\omega_{q})}}\bigg)\\
&& + (\textnormal{terms with} \,  \omega \rightarrow -\omega)\bigg]. 
\label{mitmem}
\end{eqnarray} 

For $\omega > \omega_{D}$,
\begin{eqnarray} \nonumber
\lim_{\beta\to\infty} \frac{e^{\beta\omega}-1}{e^{\beta(\omega-\omega_{q})}-1}  \log \left(\frac{1+e^{-\beta(\Delta-\omega+\omega_{q})}}{1+e^{-\beta(\Delta+\omega-\omega_{q})}}\right) \\
= \begin{cases}
\omega-\omega_{q}-\Delta &\text{if $\Delta < (\omega-\omega_{q})$}\\
0 &\text{if $\Delta > (\omega-\omega_{q})$}
\end{cases}
\end{eqnarray} 
Using the above equation and changing variable $\omega_{q}$ to $\Omega$ as done in eqn.\ref{memallen}), the eqn.(\ref{mitmem}) in terms of phonon spectral function with replacement of upper limit of integration (as discussed in earlier case) can be written as
\begin{equation}
\frac{1}{\tau(\omega, T)} = \frac{2\pi}{\omega} \int_{0}^{\omega} d\Omega \alpha^{2}F(\Omega) (\omega-\Omega-\Delta).
\label{memgap}
\end{equation}
This expression is equivalent to the expression derived by the MFF.
\section{Conclusion}
\label{sec:conclusion}
We have discussed the frequency and temperature dependent scattering rate using different approaches. A systematic description of all the approaches with their interrelations is is presented. Further using these, the results are presented for different models for the PDOS and EDOS. The comparison study is further divided into two parts, one with the temperature independent formalisms (AF, MFF and MemF) and another for the temperature dependent formalisms (SDMF, SCF and MemF). We find that the results by memory function approach agrees quite well with Allen's approach under the zero gap and zero temperature conditions (as shown in Case-I, Sec.~\ref{sec:analytical}). The former approach is also justified for MFF (Case-II, Sec.~\ref{sec:analytical}). It is shown that the scattering rate computed by both approaches namely MFF and MemF agrees well with each other in the appropriate limit of zero temperature (as MFF is applicable only in the zero temperature limit). In the finite temperature case, the memory function result of the scattering rate also reproduce the result from SCF. In SDMF, due to the lack of gap structure in their formalism, the scattering rate show sufficient finite value at low temperature and at low frequencies. In contrast in SCF and MemF, the scattering rate is suppressed due to the presence of the gap.

From this whole analysis we conclude that the MemF approach is better choice over the other approaches to study the scattering rate. As with this approach, one can study the transport properties in all different regimes of interest. While other approaches work well in specific regimes such as the MFF is applicable for $T=0$ K case. The SCF is applicable for the high frequency regime\cite{bhalla_16a}. AF and SDMF are restricted to the systems having constant electron density of states. Hence with the MemF approach, one can get rid of these difficulties and may explain the transport properties in a more general way.

\appendix

\section{For the case of Double Lorentzian model for PDOS}
\label{app:1}
\figdoubLorentA
\figdoubLorentB
\figdoubLorentemp
In this case, we consider $\alpha^{2}F(\Omega)$ of the form of double Lorentzian as below:

\begin{eqnarray}\nonumber
 \alpha^{2}F(\Omega) &=& \frac{\Omega^{2}\Omega_{p_{1}}^{2}}{\big(\Omega^{2} - \Omega_{E_{1}}^{2}\big)^{2}+(\Gamma_{1} \Omega)^{2}} \\
 && + \frac{\Omega^{2}\Omega_{p_{2}}^{2}}{\big(\Omega^{2} - \Omega_{E_{2}}^{2}\big)^{2}+\big(\Gamma_{2} \Omega\big)^{2}}.
\end{eqnarray}
Here $\Omega_{p_{1}}$, $\Omega_{p_{2}}$ are the weighted factors which we kept fixed at values $0.01$ and $0.05$ eV respectively. The other parameters such as the peak frequencies $\Omega_{1}$ and $\Omega_{2}$ both vary as $0.05$, $0.1$ and $0.2$ eV and both the Lorentzian widths $\Gamma_{1}$ and $\Gamma_{2}$ as $0.01$ and $0.1$ eV. Here again we use the same model for EDOS as in the previous cases. In Fig.~\ref{fig:doubLorentA}, we kept fixed the values corresponds to the second Lorentzian and vary the parameters of first Lorentzian. We observe that the scattering rate show kinks in the magnitude due to the two peak frequencies. In addition to this, we find that the other features remain same as the earlier cases. While in the other case where we vary the parameter values of second Lorentzian, we find that there is only one kink in the scattering rate behavior due to the phonon spectral function as shown in Fig.~\ref{fig:doubLorentB}. Here there is a suppression of the behavior by first Lorentzian due to the small value of the first weighted factor.

In Fig.~\ref{fig:doubLorentemp}, the temperature dependent behavior of the scattering rates using SDMF, SCF and MemF are plotted by changing temperature values and keeping other values fixed. We observe that the scattering rate show first kink around $0.05$ eV and then second around $0.2$ eV as shown in Fig.~\ref{fig:doubLorentemp4} for the case of $T = 300$ K.

\end{document}